\documentclass[twocolumn,english,aps,prd,reprint,groupedaddress,nobibnotes,notitlepage,nofootinbib,preprintnumbers,showpacs]{revtex4-1}
\pdfoutput=1
\usepackage{lmodern}

\usepackage[T1]{fontenc}
\usepackage[latin9]{inputenc}
\usepackage{geometry}
\usepackage{color}
\usepackage{babel}
\usepackage{amsmath}
\usepackage{amssymb}
\usepackage{graphicx}
\usepackage{esint}
\usepackage[unicode=true,pdfusetitle,
 bookmarks=true,bookmarksnumbered=false,bookmarksopen=false,
 breaklinks=false,pdfborder={0 0 1},backref=false,colorlinks=true]
 {hyperref}
\hypersetup{
 citecolor=blue,linkcolor=blue,urlcolor=blue}

\makeatletter

\newcommand{\lyxdot}{.}

 \@ifundefined{textcolor}{}
 {%
   \definecolor{BLACK}{gray}{0}
   \definecolor{WHITE}{gray}{1}
   \definecolor{RED}{rgb}{1,0,0}
   \definecolor{GREEN}{rgb}{0,1,0}
   \definecolor{BLUE}{rgb}{0,0,1}
   \definecolor{CYAN}{cmyk}{1,0,0,0}
   \definecolor{MAGENTA}{cmyk}{0,1,0,0}
   \definecolor{YELLOW}{cmyk}{0,0,1,0}
 }

\usepackage{babel}

\makeatother

\begin{document}

\preprint{\hbox{CALT-68-2853} }

\title{Attractor solutions in scalar-field cosmology}

\author{Grant N. Remmen}

\email{gremmen@theory.caltech.edu}

\author{Sean M. Carroll}

\email{seancarroll@gmail.com}

\affiliation{Division of Physics, Mathematics \& Astronomy, California Institute
of Technology}

\date{\today}

\begin{abstract}
Models of cosmological scalar fields often feature ``attractor solutions''
to which the system evolves for a wide range of initial conditions.
There is some tension between this well-known fact and another well-known
fact: Liouville's theorem forbids true attractor behavior in a Hamiltonian
system. In universes with vanishing spatial curvature, the field variables
$\phi$ and $\dot{\phi}$ specify the system completely, defining
an effective phase space. We investigate whether one can define a
unique conserved measure on this effective phase space, showing that
it exists for $m^{2}\phi^{2}$ potentials and deriving conditions
for its existence in more general theories. We show that apparent
attractors are places where this conserved measure diverges in the
$\phi$-$\dot{\phi}$ variables and suggest a physical understanding
of attractor behavior that is compatible with Liouville's theorem. 
\end{abstract}
\maketitle

\section{Introduction}

Two of the favorite moves in the repertoire of the modern theoretical
cosmologist are (1) positing one or more scalar fields whose energy
density exerts an important influence on the evolution of the universe
and (2) claiming (or at least aspiring to be able to claim) that certain
conditions or behaviors qualify as ``natural''. These tendencies
meet in the notion of cosmological attractors: dynamical conditions
under which evolving scalar fields approach a certain kind of behavior
without finely-tuned initial conditions \cite{tachyon,Urena,Belinsky,Piran,Peebles,LiddleLyth,Liddle,Kiselev,Ferreira,Downes,Steinhardt,LyapAtr},
whether in inflationary cosmology or late-time quintessence models.
In dynamical systems theory, attractor behavior describes situations
where a collection of phase-space points evolve into a certain region
and never leave. This is incompatible with Liouville's theorem, which
states that the volume of a region of phase space is invariant under
time evolution. Hamiltonian systems, of which scalar-field cosmologies
(Einstein's equation plus a dynamical scalar field, restricted to
homogeneous configurations) are examples, obey Liouville's theorem,
and therefore cannot support true attractor behavior.

So what is going on? In this paper we reconcile the appearance of
attractor solutions in scalar-field cosmologies with their apparent
mathematical impossibility by making two points. First, we point out
the fact (well-known, although rarely stated explicitly) that the
combined gravity/scalar-field equations exhibit an apparently accidental
simplification in the case of flat universes. This simplification
allows us to express the complete evolution in terms of an effective
two-dimensional ``phase space'' with coordinates $\phi$ and $\dot{\phi}$,
even though the true phase space is four-dimensional (since the scale
factor and its conjugate momentum are independent variables). Of course,
$\phi$ and $\dot{\phi}$ aren't canonical coordinates on phase space,
so the measure ${\rm d}\dot{\phi}\wedge{\rm d}\phi$ isn't very physically
meaningful.

Our second point is that it is seemingly possible to define a conserved
measure on the $\phi$-$\dot{\phi}$ effective phase space, although
this measure looks very different from ${\rm d}\dot{\phi}\wedge{\rm d}\phi$.
We cannot rigorously prove its existence in general, but we can show
that it corresponds to a Lagrangian on effective phase space if it
does exist; in the simple example of a canonical scalar field with
a quadratic potential, we show that a unique measure on effective
phase space exists and derive some of its properties. By construction,
there can be no ``attractor'' solutions with respect to this measure.
Nevertheless, we suggest there is a relevant sense in which attractor
solutions are physically meaningful, if certain functions of the phase-space
variables are directly observable. Finally, we comment on the connection
between this classical analysis and the boundary induced on phase
space by the Planck scale.

\section{Phase Space, Measures, and Attractors\label{MathPrelims}}

We start by reviewing scalar-field cosmology in phase space, following
\citet*{GHS} (GHS). There are subtleties due to the fact that GR
is a constrained system. In this section, we also discuss the intuitive
idea of an attractor and contrast it with Hamiltonian behavior.

Because a phase space $\Gamma$ of dimension $2n$ is a symplectic
manifold, there is a closed two-form defined on $\Gamma$, 
\begin{equation}
\omega=\sum_{i=1}^{n}\mathrm{d}p_{i}\wedge\mathrm{d}q^{i}.
\end{equation}
This symplectic form defines the Liouville measure, 
\begin{equation}
\Omega=\frac{\left(-1\right)^{n\left(n-1\right)/2}}{n!}\omega^{n}.
\end{equation}
Liouville's theorem from classical mechanics states that this measure
is conserved along the Hamiltonian flow vector $X_{\mathcal{H}}$.
That is, given trajectories that initially cover some region $S\subset\Gamma$
and that evolve under $X_{\mathcal{H}}$ to cover region $S^{\prime}$,
we have 
\begin{equation}
\int_{S}\Omega=\int_{S^{\prime}}\Omega.
\end{equation}
Equivalently, the Lie derivative of $\Omega$ vanishes along $X_{\mathcal{H}}$,
\begin{equation}
\pounds_{X_{\mathcal{H}}}\Omega=0.
\end{equation}

Because the metric component $g_{00}$ is not a propagating degree
of freedom in the Einstein--Hilbert action, general relativity is
a constrained system, in which the Hamiltonian $\mathcal{H}$ is set
to a boundary-condition-dependent constant along physical trajectories.
That is, trajectories are confined to a hypersurface in $\Gamma$
of dimension $2n-1$ for which $\mathcal{H}=\mathcal{H}_{\star}$;
we will call this the Hamiltonian constraint surface, 
\begin{equation}
C=\Gamma/\left\{ \mathcal{H}=\mathcal{H}_{\star}\right\} .
\end{equation}
The Hamiltonian flow vector, describing Hamiltonian evolution of trajectories
in $C$, is 
\begin{equation}
X_{\mathcal{H}}=\frac{\partial\mathcal{H}}{\partial p_{i}}\frac{\partial}{\partial q^{i}}-\frac{\partial\mathcal{H}}{\partial q^{i}}\frac{\partial}{\partial p_{i}},\label{Hamflowvector}
\end{equation}
where $(q^{i},p_{i})$ are the canonical coordinates and their conjugate
momenta. The space of trajectories (as opposed to states) can be defined
by taking the quotient 
\begin{equation}
M=C/X_{\mathcal{H}}.
\end{equation}

Previously, \citet{GHS} contructed the unique measure on $M$ for
FRW universes. The GHS measure is unique in that it is positive, independent
of parametrization, and respects the symmetries of the problem without
introducing additional structures. It is obtained from the symplectic
form $\omega$ by identifying the $n$th coordinate of phase space
$\Gamma$ as time $t$, so that 
\begin{equation}
\omega=\tilde{\omega}+\mathrm{d}\mathcal{H}\wedge\mathrm{d}t=\sum_{i=1}^{n-1}\mathrm{d}p_{i}\wedge\mathrm{d}q^{i}+\mathrm{d}\mathcal{H}\wedge\mathrm{d}t.
\end{equation}
The corresponding measure, a $\left(2n-2\right)$-form, is 
\begin{equation}
\Theta=\frac{\left(-1\right)^{\left(n-1\right)\left(n-2\right)/2}}{\left(n-1\right)!}\tilde{\omega}^{n-1}.
\end{equation}

The metric describing an FRW universe is 
\begin{equation}
\mathrm{d}s^{2}=-N^{2}\mathrm{d}t^{2}+a^{2}\left(t\right)\left(\frac{\mathrm{d}r^{2}}{1-\kappa r^{2}}+r^{2}\mathrm{d}\Omega^{2}\right),
\end{equation}
where $a\left(t\right)$ is the scale factor, normalized to unity
at some time $t_{0}$, and $N$ is the lapse function. The curvature
parameter $\kappa\in\mathbb{R}$ has dimensions of $\left[\mathrm{length}\right]^{-2}$.
We may also define $k=\kappa R_{0}^{2}\in\left\{ 0,\pm1\right\} $,
where $R_{0}$ is the radius of curvature of the universe when $a(t_{0})=1$.
\begin{figure*}
\noindent \begin{centering}
\includegraphics[width=3.375in]{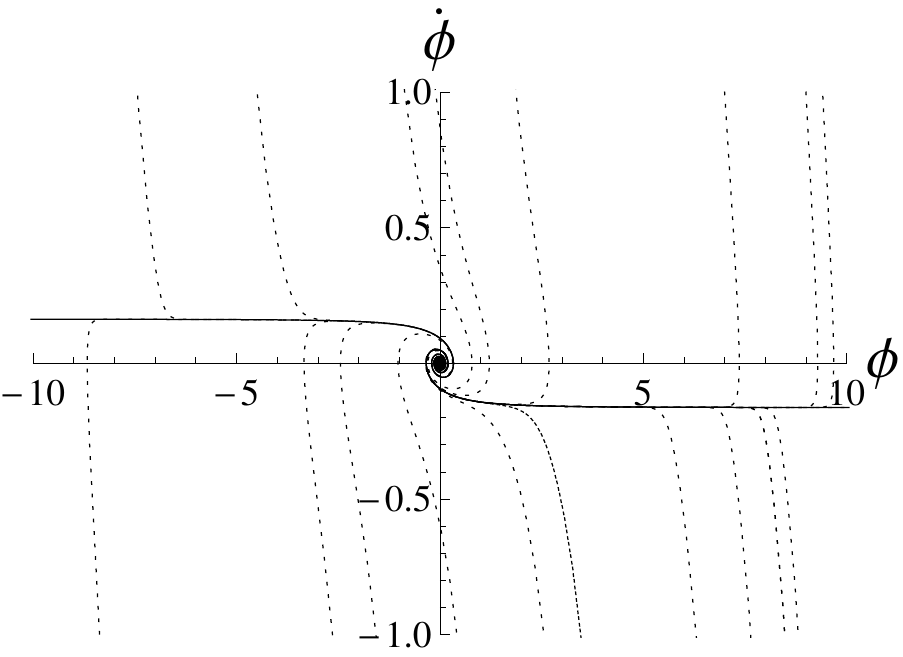}$\;\;\;\;\;$\includegraphics[width=3.375in]{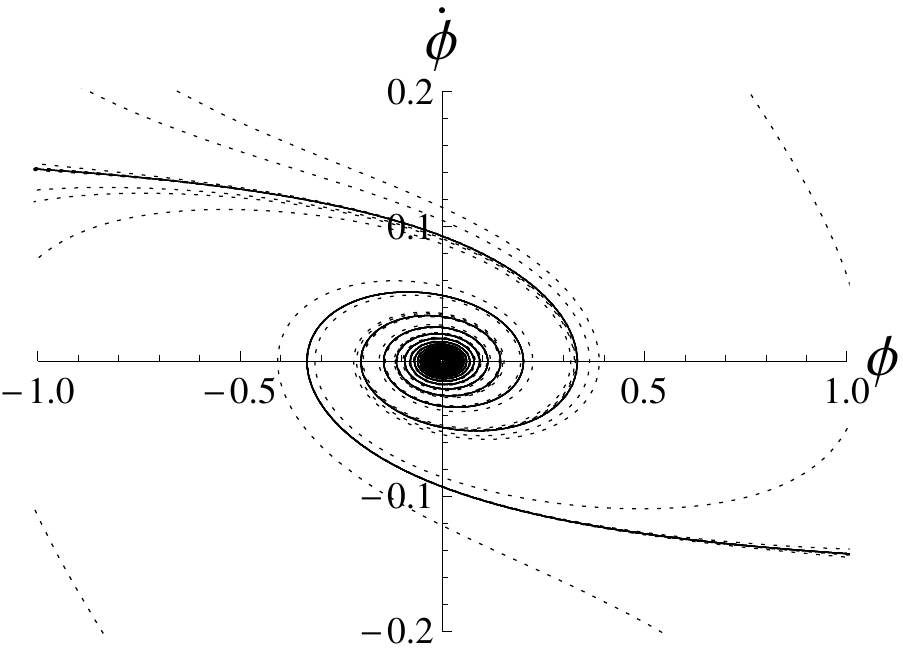} 
\par\end{centering}

\caption{\label{attractors}Apparent attractor solutions for an $m^{2}\phi^{2}$
potential, with equation of motion $\ddot{\phi}+\sqrt{3/2}\sqrt{m^{2}\phi^{2}+\dot{\phi}^{2}}\dot{\phi}+m^{2}\phi=0$.
Solid: the apparent attractors; dotted: numerical solutions for random
initial conditions. Plots are in $\phi$-$\dot{\phi}$ space, in units
where $M_{\mathrm{Pl}}=1$; the scalar mass is chosen to be $m=0.2M_{\mathrm{Pl}}$.
At large field values, the solutions are approximated by the lines
$\dot{\phi}=\pm\sqrt{2/3}m$, while for small field values, all solutions
converge on the origin.}
\end{figure*}

\begin{figure*}
\noindent \begin{centering}
\includegraphics[width=3.375in]{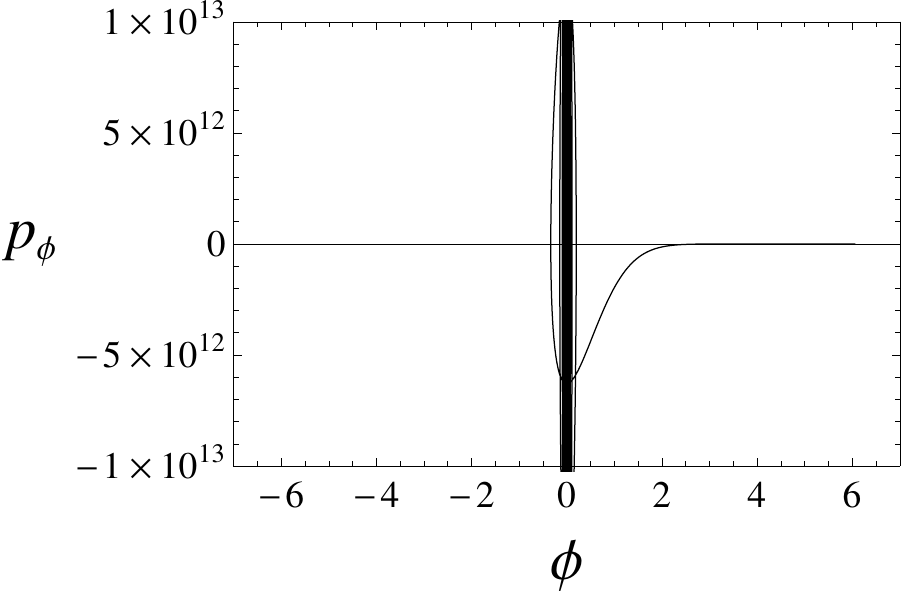}$\;\;\;\;\;$\includegraphics[width=3.375in]{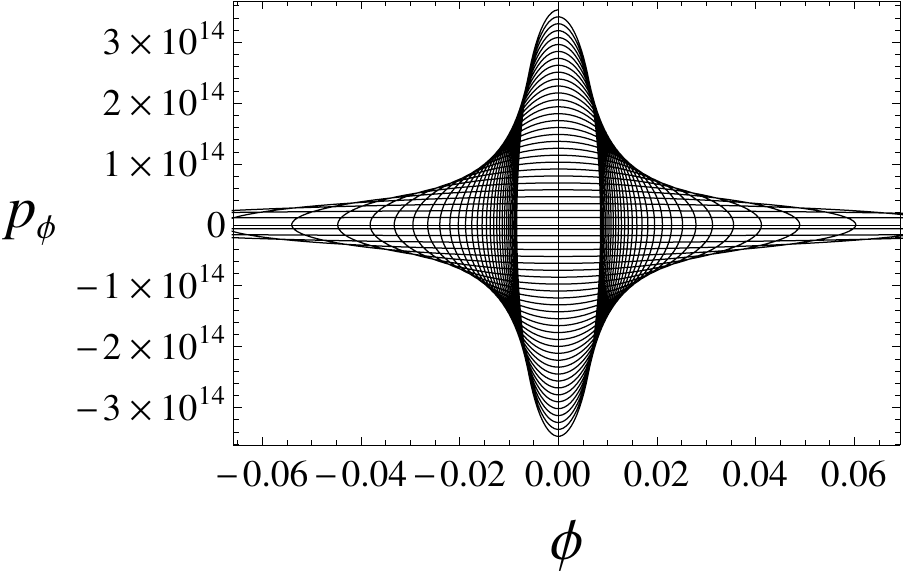} 
\par\end{centering}

\caption{\label{phipphiplot}Numerical solution for evolution of an FRW universe
with an $m^{2}\phi^{2}$ potential, with initial conditions $(\phi,\dot{\phi})=(6,0.25)$
at $a\left(t=0\right)=1$, plotted in $(\phi,p_{\phi})$ coordinates,
where $p_{\phi}=a^{3}\dot{\phi}$ is the canonical momentum conjugate
to $\phi$. Units are chosen such that $M_{\mathrm{Pl}}=1$, with
the scalar mass $m=0.2M_{\mathrm{Pl}}$. The apparent attractor behavior
seen in Fig. \ref{attractors} disappears in these coordinates.}
\end{figure*}

Studying the dynamics of the FRW scale factor coupled to some matter
source is known as the minisuperspace approximation. The minisuperspace
Lagrangian for gravity plus a scalar field with potential $V\left(\phi\right)$
is 
\begin{equation}
\mathcal{L}=3M_{\mathrm{Pl}}^{2}\left(Na\kappa-\frac{a\dot{a}^{2}}{N}\right)+a^{3}\left[\frac{\dot{\phi}^{2}}{2N}-NV\left(\phi\right)\right].
\end{equation}
The canonical momenta, defined as $p_{i}=\partial\mathcal{L}/\partial\dot{q}^{i}$,
are 
\begin{equation}
p_{N}=0,\; p_{a}=-6N^{-1}M_{\mathrm{Pl}}^{2}a\dot{a},\;\mathrm{and}\;\; p_{\phi}=N^{-1}a^{3}\dot{\phi}.\label{momenta}
\end{equation}
Note that $N$ is a Lagrange multiplier: it is non-dynamical and will
not be a part of the phase space. Performing a Legendre transformation,
the Hamiltonian, in units where $M_{\mathrm{Pl}}=\sqrt{\hbar c/8\pi G}=1$,
is 
\begin{equation}
\begin{aligned}\mathcal{H} & =N\left[-\frac{p_{a}^{2}}{12a}+\frac{p_{\phi}^{2}}{2a^{3}}+a^{3}V\left(\phi\right)-3a\kappa\right]\\
 & =-3a^{3}N\left\{ \left(\frac{\dot{a}}{a}\right)^{2}+\frac{\kappa}{a^{2}}-\frac{1}{3}\left[\frac{1}{2}\dot{\phi}^{2}+V\left(\phi\right)\right]\right\} .
\end{aligned}
\label{Hamiltonian}
\end{equation}
The equation of motion for $N$ sets it equal to an arbitrary constant,
which we choose to be unity henceforth. Varying the action with respect
to $N$ gives the Hamiltonian constraint for FRW universes, $\mathcal{H}_{\star}=0$,
which is equivalent to the Friedmann equation, 
\begin{equation}
H^{2}=\frac{1}{3}\left[\frac{1}{2}\dot{\phi}^{2}+V\left(\phi\right)\right]-\frac{\kappa}{a^{2}},\label{FriedmannEquation}
\end{equation}
where the Hubble parameter is $H\equiv\dot{a}/a$. Thus, $\Gamma$
is four-dimensional, with $(\phi,\; p_{\phi},\; a,\; p_{a})$ being
a possible parametrization. The Hamiltonian constraint surface $C$,
once a value of $\kappa$ is chosen, is three-dimensional. The space
of trajectories $M$ is two-dimensional. The GHS measure can be written
as the Liouville measure with the Hamiltonian constraint: 
\begin{equation}
\Theta=\left.\left(\mathrm{d}p_{a}\wedge\mathrm{d}a+\mathrm{d}p_{\phi}\wedge\mathrm{d}\phi\right)\right|_{\mathcal{H}=0}.\label{GHSmeasure}
\end{equation}

A true attractor in phase space can be thought of as a region toward
which phase-space trajectories converge when plotted in canonical
coordinates. More formally, an attractor is defined \cite{Milnor}
as a region $A\subset\Gamma$ with the following properties: 
\begin{enumerate}
\item $A$ is compact; 
\item Given a trajectory $\{\mathcal{P}(t,x_{0})\}\subset\Gamma$ beginning
at $\mathcal{P}(t_{0},x_{0})=x_{0}\in A$, $\mathcal{P}(t,x_{0})\in A$
for all $t>t_{0}$; 
\item There exists a basin of attraction, a neighborhood $B$ of $A$ such
that for all $x_{B}\in B$ and for any neighborhood $N$ of $A$,
there exists $t_{N}$ such that $\mathcal{P}(t,x_{B})\in N$ for all
$t>t_{N}$; 
\item Properties 2. and 3. are not satisfied by any $A^{\prime}\subsetneq A$. 
\end{enumerate}
There are other, related, definitions of attractors in the mathematical
literature \cite{Auslander}; in particular, a definition in terms
of Lyapunov stability is possible (cf. Sec. \ref{PhysicalMeaning},
below).

An immediate consequence of Liouville's theorem is that no true attractor
can exist in the phase space of a system described by a Hamiltonian
\cite{Gibbs}; see also Ref. \cite{MTW}, Sec. 22.6. Intuitively,
if a bundle of trajectories converges along a particular axis in phase
space in a given coordinate system, it must compensatingly spread
out along other axes, to conserve the total phase-space measure. Though
we may wish to describe such behavior as an ``attractor'', it is
always possible to remove this apparent convergence by a canonical
change of coordinates: in essence, there is no coordinate-independent
notion of an attractor in the full four-dimensional phase space describing
scalar-field cosmology in an FRW universe \cite{ChienYao}.

Despite the fact that it does not rigorously exist, however, the intuitive
idea of an attractor appears in the literature on scalar field cosmology,
though a definition of what is meant by an ``attractor'' is often
left implicit. This often occurs as a result of plotting trajectories
in some non-canonical phase-space variables, most commonly $\phi$
versus $\dot{\phi}$ \cite{tachyon,Urena,Downes}. However, as one
can see in Figs. \ref{attractors} and \ref{phipphiplot}, apparent
attractor behavior in $(\phi,\dot{\phi})$ coordinates need not correspond
to attractor behavior when plotted in $(\phi,p_{\phi})$. Furthermore,
recall that the full phase space $\Gamma$ is four-dimensional, not
two-dimensional: $a$ and $p_{a}$ are suppressed in Figs. \ref{attractors}
and \ref{phipphiplot}, and initially nearby trajectories would generally
spread in these variables. In other papers, the notion of an ``attractor''
is used in a manifestly coordinate-dependent manner, with respect
to some physical observables that either become smaller with time
\cite{Steinhardt} or for which differences between initially different
trajectories vanish rapidly in some particular coordinates \cite{LiddleLyth,Liddle,Kiselev}.

It is easy to see why such behavior is described as attractor-like:
one simply looks at the plots, perhaps implicitly assuming a ``graph
paper measure'' $\mathrm{d}\dot{\phi}\wedge\mathrm{d}\phi$. Though
this assumption seems natural, it is a coordinate-dependent artifact,
as $\phi$ and $\dot{\phi}$ are not canonically conjugate. It is
our aim in this work to make all of these notions more rigorous, examining
both the issue of the measure on the space of field variables and
the definition of apparent attractor-like behavior. Our results should
help to create a common, more mathematically valid, and less \emph{ad
hoc} language for comparing results between different models of scalar-field
cosmology.

\section{Effective Phase Space for a Single Scalar Field\label{EffPhSp}}

In this section, we identify a sense in which $\phi$ and $\dot{\phi}$,
though not canonically conjugate, are special coordinates for universes
with zero spatial curvature. That is, we will show that the full four-dimensional
phase space $\Gamma$ is larger than necessary to fully capture the
dynamics of scalar field cosmology for flat universes; $\phi$-$\dot{\phi}$
space can be regarded as an effective phase space, in a sense that
will be made precise. We proceed first by defining the notion of a
vector field invariant map, making as general and coordinate-independent
definitions as possible. In essence, given a map between two manifolds
and a vector field on the first manifold, the map is vector field
invariant if it provides a way of uniquely specifying a vector field
on the second manifold. We will find that the map from $\Gamma$ to
$\phi$-$\dot{\phi}$ space for flat universes is vector field invariant
with respect to the Hamiltonian flow vector.

\subsection{Vector field invariant maps}

Before investigating whether there is a sense in which the non-canonical
coordinates $(\phi,\dot{\phi})$ constitute a parametrization with
any special mathematical properties, we first require some definitions
and notation. Given two manifolds $M$ and $N$, a mapping $\psi:\, M\rightarrow N$,
and $f\in\mathcal{F}\left(N\right)$, where $\mathcal{F}\left(N\right)$
is the space of smooth real-valued functions with domain $N$, the
pullback $\psi^{\star}:\,\mathcal{F}\left(N\right)\rightarrow\mathcal{F}\left(M\right)$
of $f$ by $\psi$ is defined by: 
\begin{equation}
\psi^{*}f=\left(f\circ\psi\right):\, M\rightarrow\mathbb{R}.
\end{equation}
We can think of $f$ as specifying a coordinate on $N$ and the pullback
as specifying a coordinate on $M$. Now, at a given point $p\in M$,
we may regard a vector $X\left(p\right)$ as a function $X_{p}:\,\mathcal{F}\left(M\right)\rightarrow\mathbb{R}$.
If we think of $g\in\mathcal{F}\left(M\right)$ as specifying a coordinate
(also called $g$) on $M$, then $X_{p}\left(g\right)$ gives the
value of the $g$-component of the vector at $p$. A vector field
$X$ on $M$ is the assignment of a vector $X_{p}$ to each point
$p\in M$ in a continuous and smooth fashion. Given the map $\psi:\, M\rightarrow N$
and a function $f:\, N\rightarrow\mathbb{R}$, the pushforward of
$X$ at $\psi\left(p\right)\in N$ is 
\begin{equation}
\left(\psi_{*}X\right)_{\psi\left(p\right)}\left(f\right)=X_{p}\left(\psi^{*}f\right).
\end{equation}
We note that the pushforward $\psi_{*}$ is a map from the tangent
space at $p$, $T_{p}M$, into $T_{\psi(p)}N$ and that $\psi_{*}X:\,\mathcal{F}\left(N\right)\rightarrow\mathbb{R}$.
In this sense, we can write $\psi_{*}\left(X\right)=X\circ\psi^{*}$.
For further reference, see Appendix C of \cite{Wald} and Appendix
A of \cite{SeanGR}.

We may now define ``vector field invariance'', a way of formalizing
the idea that a many-to-one map creates a unique vector field. Suppose
we have a map $\psi:\, M\rightarrow N$ and vector field $X$ on $M$.
For each point $q\in N$, write the preimage in $M$ as $\psi^{-1}\left(q\right)=\{p\in M\;|\;\psi\left(p\right)=q\}$.
Then say that the map $\psi$ \emph{is vector field invariant with
respect to $X$} if for any function $f\in\mathcal{F}\left(N\right)$
and for all $q\in N$, $X_{p}(\psi^{*}f)=X_{p^{\prime}}(\psi^{*}f)$
for all $p,\, p^{\prime}\in\psi^{-1}\left(q\right)$. If a map $\psi$
is vector field invariant with respect to $X$, we may write $X_{p}(\psi^{*}f)=\tilde{X}_{q}(f)$
without ambiguity for $p\in\psi^{-1}\left(q\right)$. Then we have
a unique vector field $\tilde{X}$ on $N$. We can say that $\tilde{X}$
is the vector field \emph{induced by $X$ on $N$} \emph{under the
(vector field invariant) map }$\psi$.

The images of integral curves that are distinct under a vector field
invariant map do not intersect. If we have a vector field invariant
map $\psi:\, M\rightarrow N$, where $M$ has vector field $X$, then
the images of integral curves of $X$ are integral curves of $\tilde{X}$
in $N$. Therefore, by uniqueness, given two integral curves in $M$
not mapped onto each other in $N$ by $\psi$, their images in $N$
cannot intersect. If $M$ is the phase space for some Hamiltonian
system and $\psi:\, M\rightarrow N$ is vector field invariant with
respect to the Hamiltonian flow vector, one can therefore think of
$N$ as an \emph{effective phase space}.

\subsection{A map for FRW universes}

We will now show that, for scalar-field cosmology in a flat universe,
the choice of $\phi$ and $\dot{\phi}$ as coordinates allows one
to eliminate $a$ and $\dot{a}$ and thus reduce the dynamical phase
space to two dimensions. Consider a map $\chi$ from the Hamiltonian
constraint three-manifold $C$ to a two-manifold $K$, where $\chi^{-1}\left(q\right)$
is the set of all points in $C$ with equal values of $\phi$ and
$\dot{\phi}$. That is, $K$ is isomorphic to $\phi$-$\dot{\phi}$
space. We will show that in a flat universe ($\kappa=0$) with a scalar
field described by a potential $V\left(\phi\right)$ and a canonical
kinetic term, the map $\chi$ is vector field invariant with respect
to the Hamiltonian flow vector $X_{\mathcal{H}}$.

It is sufficient to exhibit one such map $\chi$, as all other maps
such that the preimage of $q\in K$ is the set of all points in $C$
with equal values of $\phi$ and $\dot{\phi}$ can be obtained from
$\chi$ via a bijection. Without loss of generality, we may therefore
specify coordinates $(\phi,\,\dot{\phi},\, a,\, H)$ on the full phase
space $\Gamma$, which are inherited by $C$, so that $C$ is parametrized
by four coordinates related by the Hamiltonian constraint. Note, however,
that none of our conclusions are dependent on choosing $a$ and $H$
as the other two coordinates. That is, the notion of vector field
invariance of $\chi:\, C\rightarrow K$ is a statement only about
$\phi$ and $\dot{\phi}$, independent of the other coordinates. With
the Hamiltonian \eqref{Hamiltonian} and flow vector \eqref{Hamflowvector},
we have 
\begin{equation}
\begin{aligned}X_{\mathcal{H}}^{\left(\phi\right)} & =\frac{p_{\phi}}{a^{3}},\\
X_{\mathcal{H}}^{\left(p_{\phi}\right)} & =-a^{3}V^{\prime}\left(\phi\right),\\
X_{\mathcal{H}}^{\left(a\right)} & =-\frac{p_{a}}{6a},\\
X_{\mathcal{H}}^{\left(p_{a}\right)} & =-\frac{p_{a}^{2}}{12a^{2}}+\frac{3p_{\phi}^{2}}{2a^{4}}-3a^{2}V\left(\phi\right)+3\kappa.
\end{aligned}
\label{vectorcomps}
\end{equation}
Using the expressions for $p_{a}$ and $p_{\phi}$ in Eq. \eqref{momenta}
to rescale and eliminate $a$ and $\dot{a}$ in favor of $\phi$ and
$\dot{\phi}$, we have 
\begin{equation}
\begin{aligned}X_{\mathcal{H}}^{(\phi)}= & \dot{\phi},\\
X_{\mathcal{H}}^{(\dot{\phi})}= & \frac{1}{a^{3}}X_{\mathcal{H}}^{(p_{\phi})}=-V^{\prime}\left(\phi\right),\\
X_{\mathcal{H}}^{(H)}= & \!-\frac{1}{6a^{2}}X_{\mathcal{H}}^{(p_{a})}\!\!=\frac{1}{2}H^{2}\!-\!\frac{1}{4}\dot{\phi}^{2}\!+\!\frac{1}{2}V\!\left(\phi\right)\!-\!\frac{\kappa}{2a^{2}}.
\end{aligned}
\end{equation}
Therefore, for $\kappa=0$, the $\phi$-, $\dot{\phi}$-, and $H$-components
of the vector field are independent of $a$. Further, from the Friedmann
equation \eqref{FriedmannEquation}, $H$ can be written as a function
of $\phi$ and $\dot{\phi}$ for $\kappa=0$. Thus, for a flat universe,
the $\phi$-, $\dot{\phi}$-, and $H$-components of the Hamiltonian
vector field $X_{\mathcal{H}}$ can be written in terms of $\phi$
and $\dot{\phi}$ alone. This is the slightly more careful version
of our previous statement that $\phi$ and $\dot{\phi}$ allow $a$
and $\dot{a}$ to be eliminated from the dynamics.\emph{ }Now, consider
the map $\chi:\, C\rightarrow K$ defined by $\chi(a,\phi,\dot{\phi},H)=(\phi,\dot{\phi})$.
Under such a map, the condition for vector field invariance for a
given vector field $X$ is simply the condition that the $\phi$-,
$\dot{\phi}$-, and $H$-components of $X$ can be written in terms
of only $\phi$ and $\dot{\phi}$. Hence, we conclude that the map
$\chi$ is vector field invariant with resepect to the Hamiltonian
vector field $X_{\mathcal{H}}$ for a flat universe.

We have shown, for a universe of zero spatial curvature, that there
is a sense in which $(\phi,\dot{\phi})$ become effective phase-space
coordinates. This formalizes the intuitive idea that the equations
of motion can be written purely in terms of these variables. There
exists a vector field invariant map with respect to the Hamiltonian
flow vector, from the full three-dimensional constraint surface $C$
to a two-dimensional manifold $K$: we find that $K$ is isomorphic
to $\phi$-$\dot{\phi}$ space. This is a nontrivial property -- it
is not in general true, given a three-dimensional surface with a Hamiltonian
vector field, that a vector field invariant mapping to a two-dimensional
manifold exists. The criterion $\kappa=0$ is necessary for $(\phi,\dot{\phi})$
to be an effective phase space; indeed, trajectories can cross in
$\phi$-$\dot{\phi}$ space if $\kappa\neq0$. Furthermore, the projection
of $\Gamma$ onto two \emph{canonical} coordinates \emph{does not}
constitute construction of an effective phase space; this fact can
be illustrated dramatically by considering orbits in $(\phi,p_{\phi})$
for an $m^{2}\phi^{2}$ potential (see Fig. \ref{phipphiplot}). In
this sense, $\phi$ and $\dot{\phi}$ are special coordinates with
which to parametrize the phase space of scalar-field cosmology.

\subsection{The geometrical picture\label{GeometricalPicture}}

One can develop more intuition about the notion of vector field invariance
by considering the geometry of the Hamiltonian constraint submanifold
embedded in the full phase space, for a specific model with $V\left(\phi\right)=m^{2}\phi^{2}/2$.
The four-dimensional phase space $\Gamma$ is foliated into three-dimensional
Hamiltonian submanifolds $C$, each with a unique value of $\kappa$,
with the Friedmann equation \eqref{FriedmannEquation} giving the
constraint.

\begin{figure*}
\noindent \begin{centering}
\includegraphics[width=1.5in]{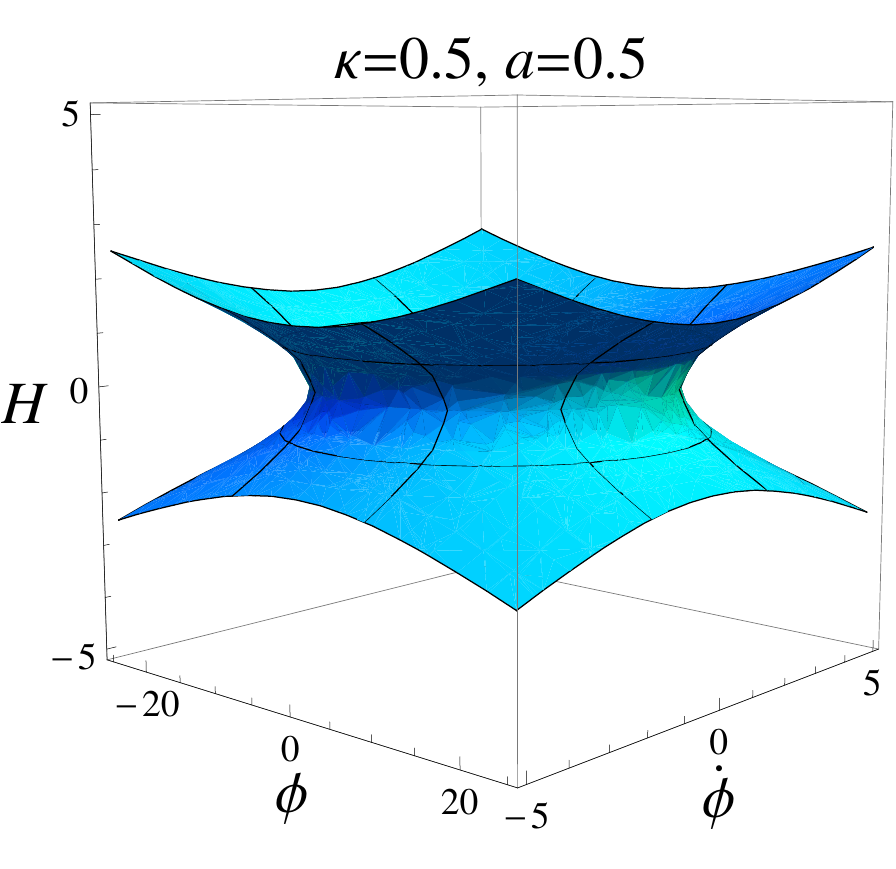}\includegraphics[width=1.5in]{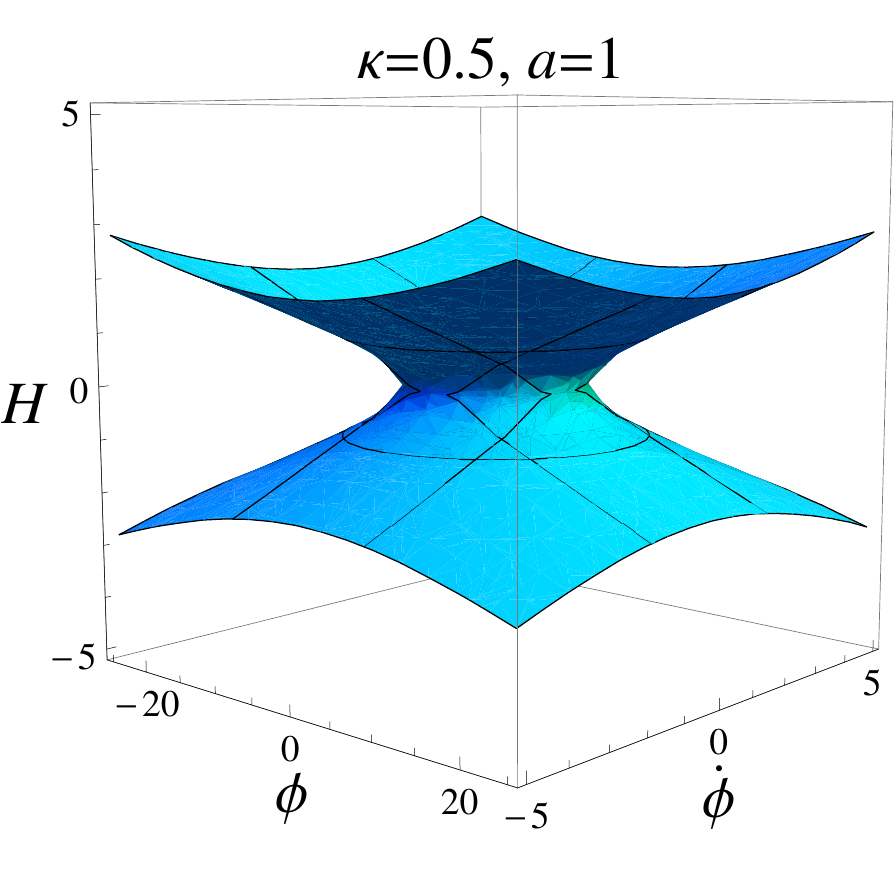}\includegraphics[width=1.5in]{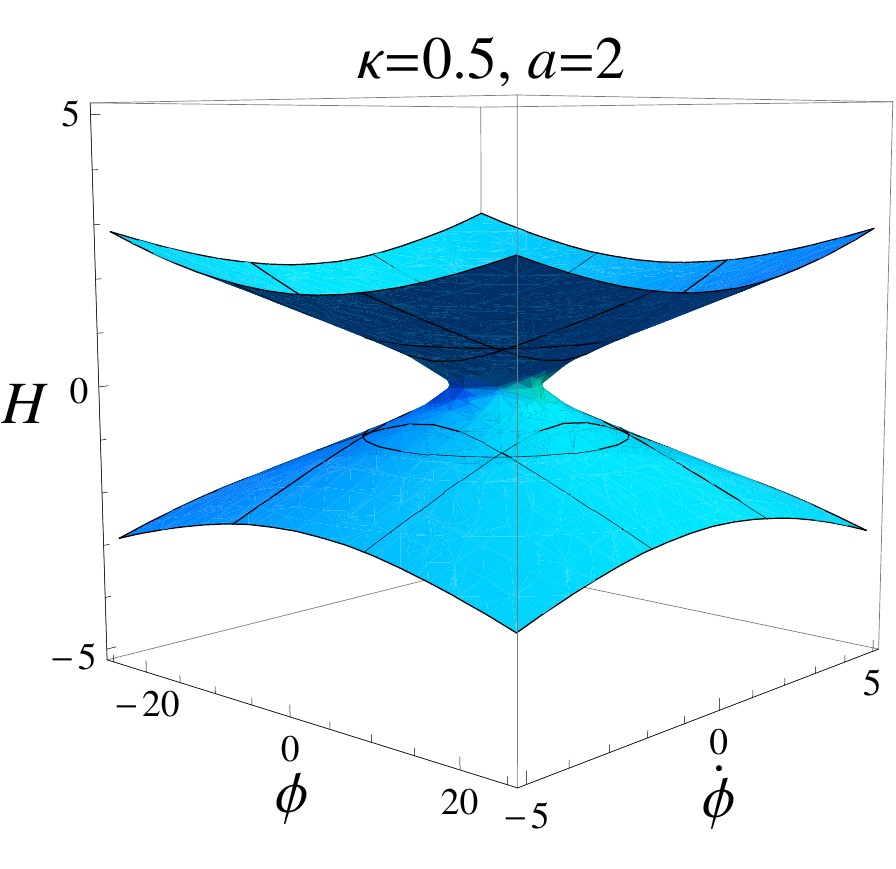}\includegraphics[width=1.5in]{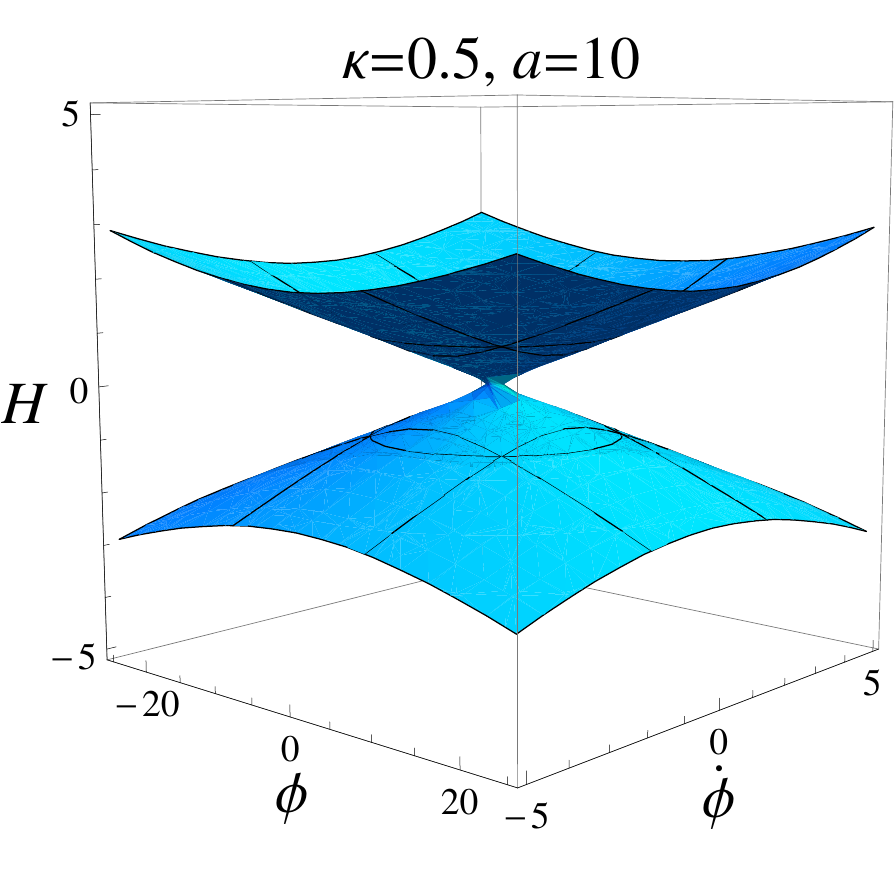} 
\par\end{centering}

\noindent \begin{centering}
\includegraphics[width=1.5in]{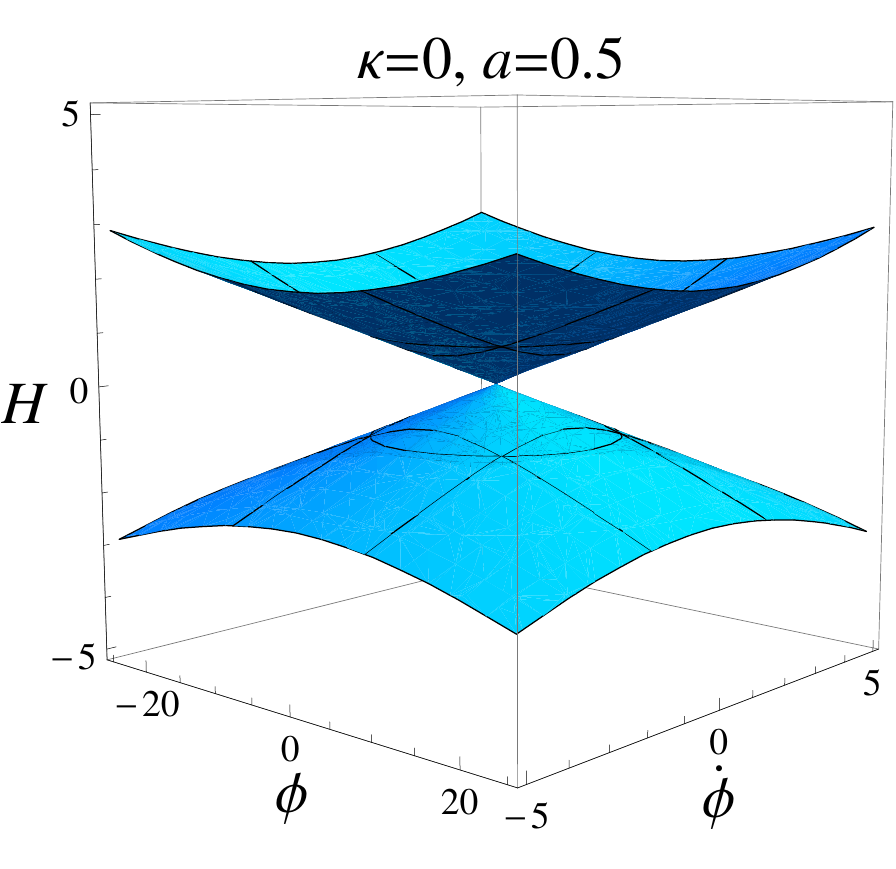}\includegraphics[width=1.5in]{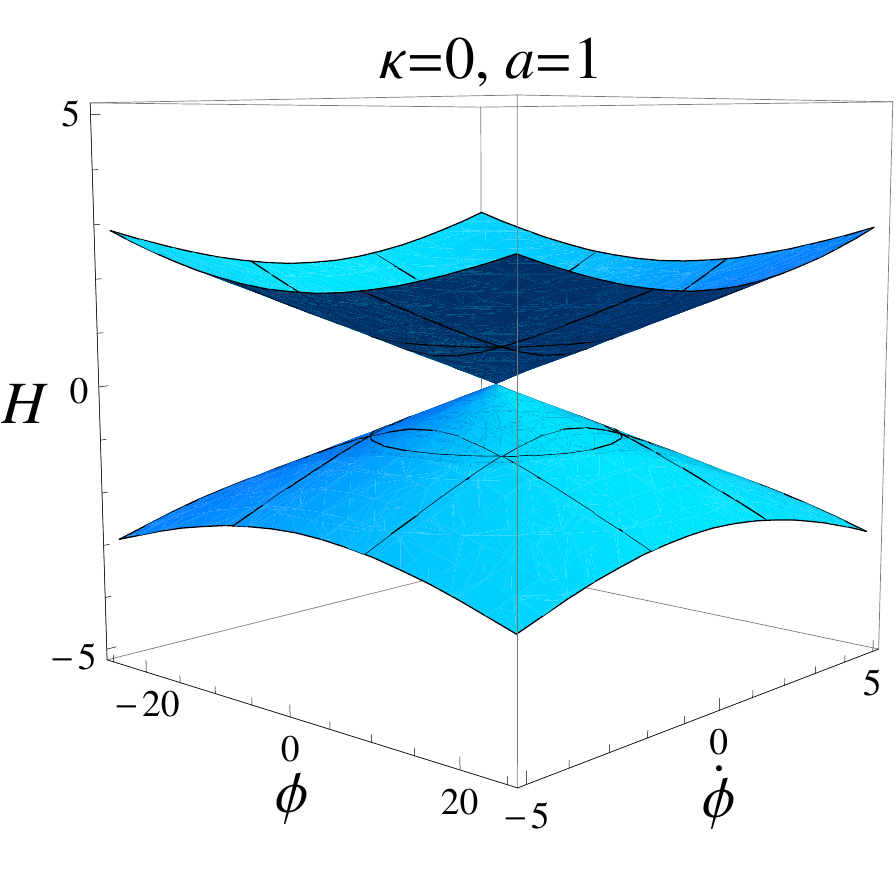}\includegraphics[width=1.5in]{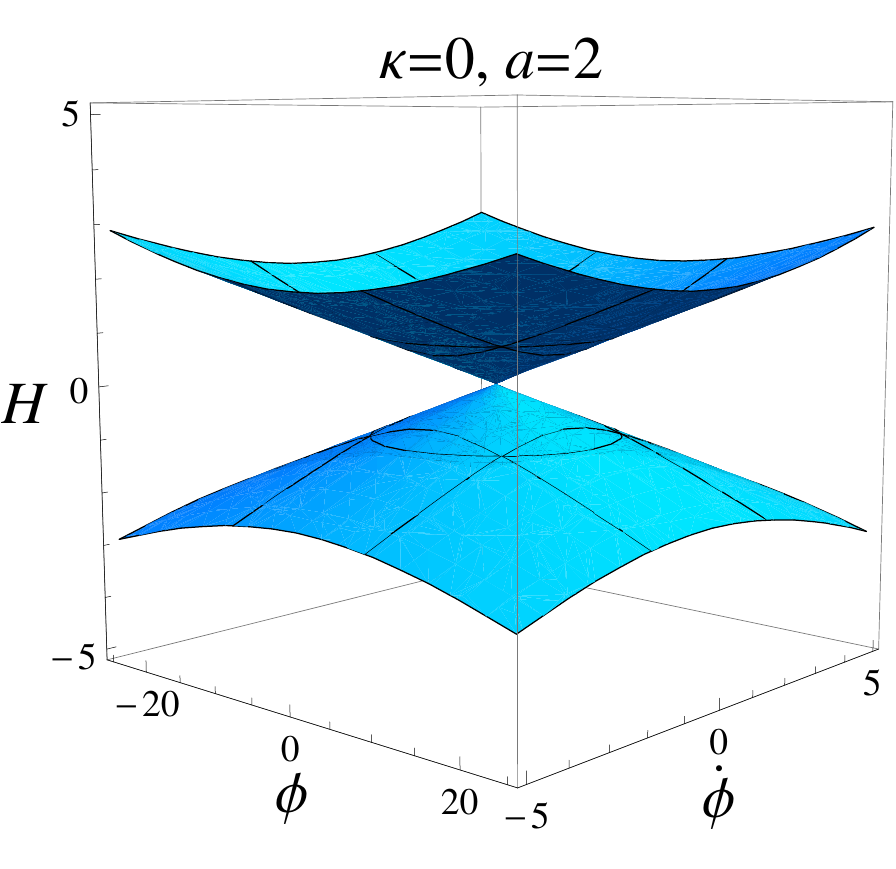}\includegraphics[width=1.5in]{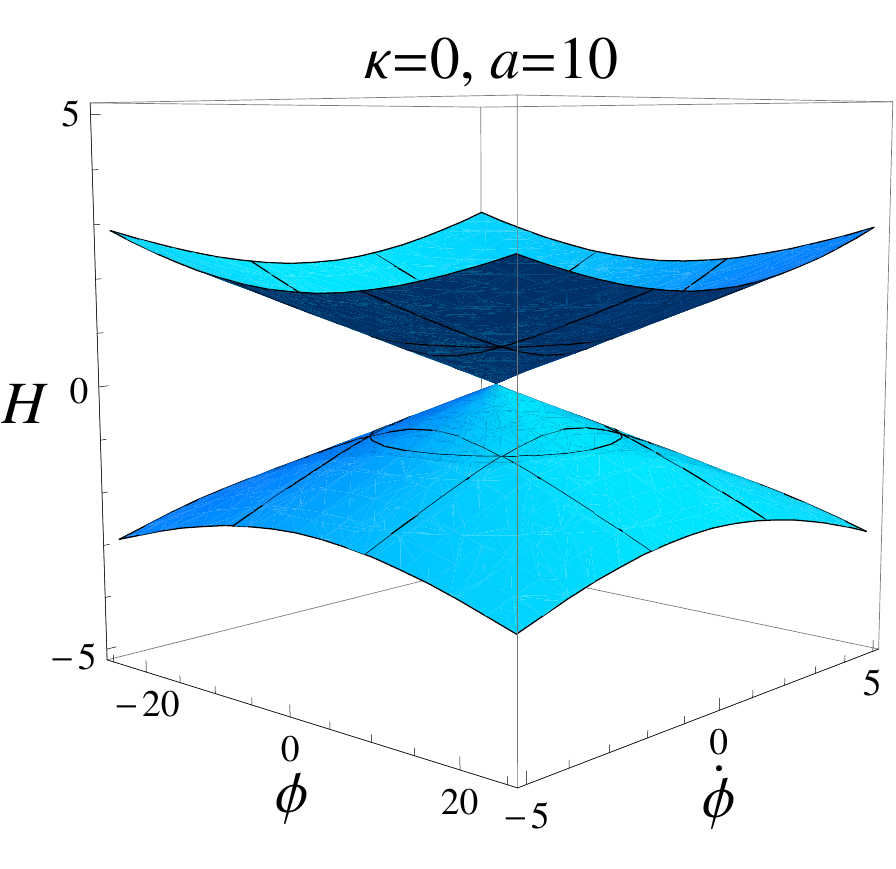} 
\par\end{centering}

\noindent \begin{centering}
\includegraphics[width=1.5in]{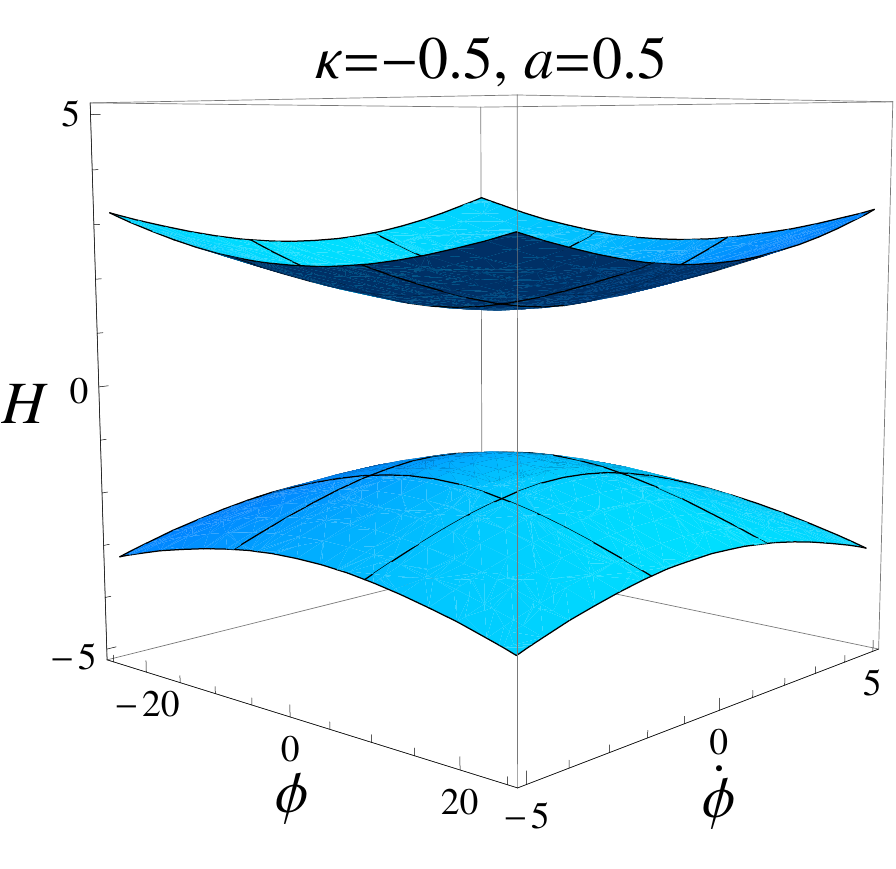}\includegraphics[width=1.5in]{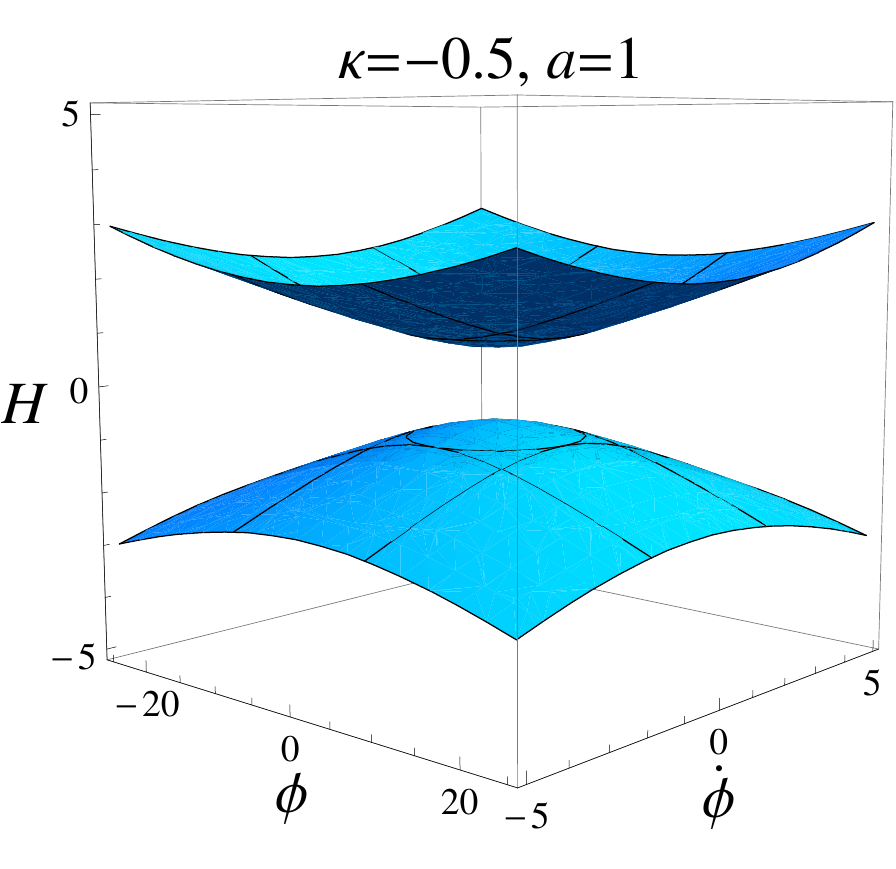}\includegraphics[width=1.5in]{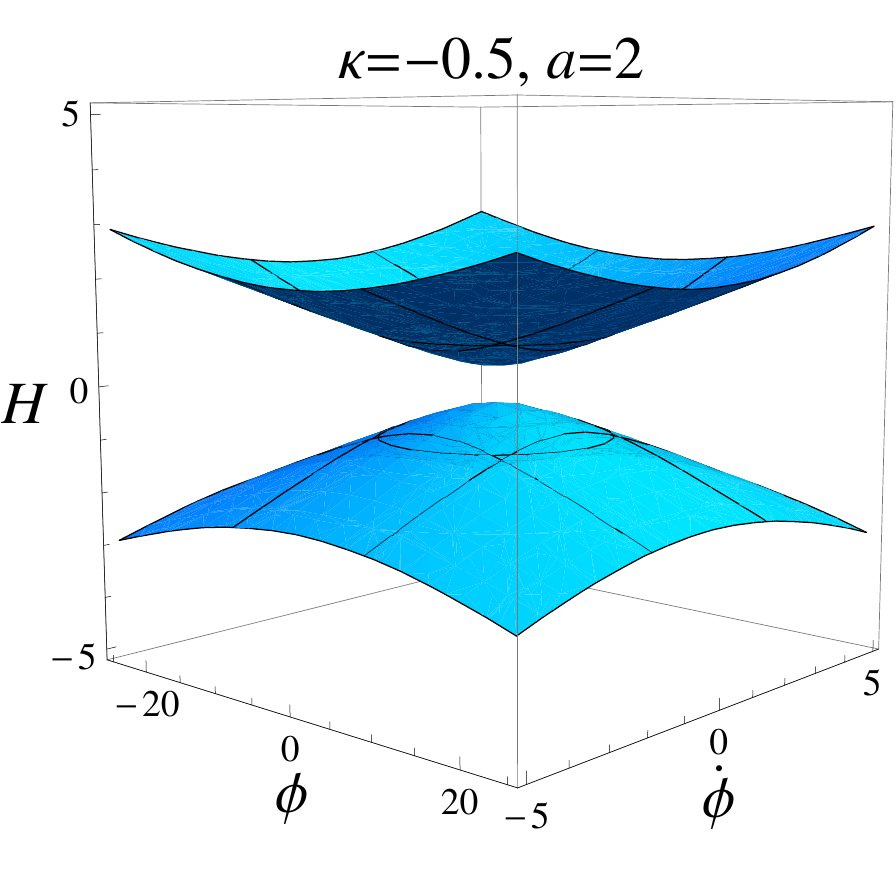}\includegraphics[width=1.5in]{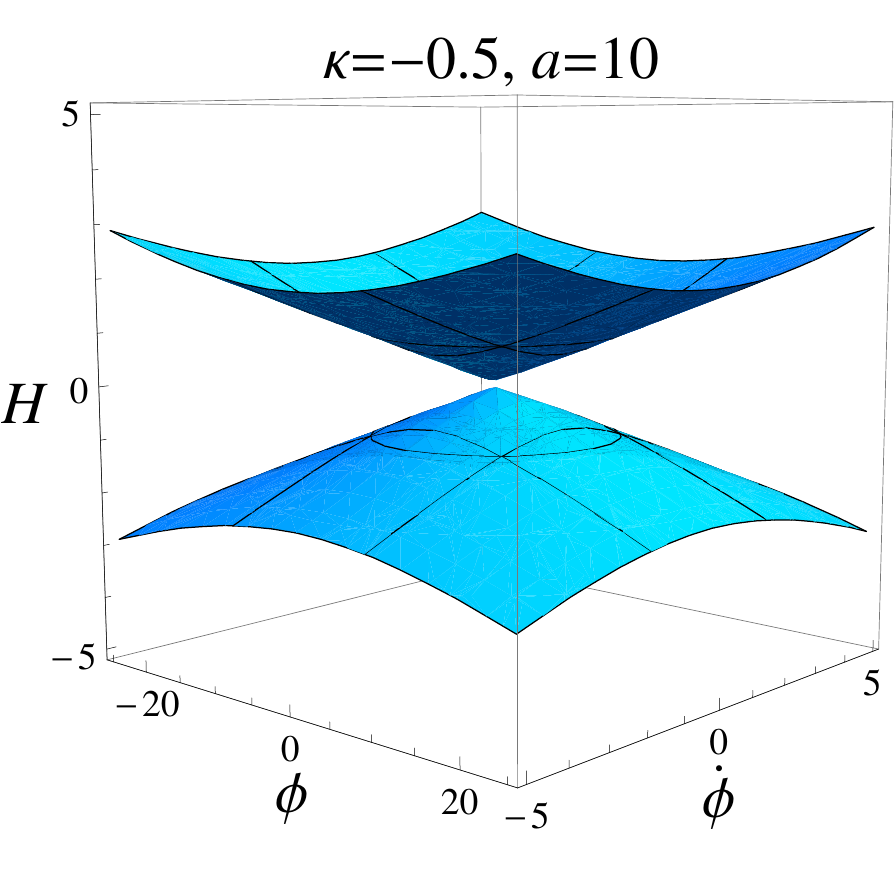} 
\par\end{centering}

\caption{\label{slices}Plots of the Hamiltonian three-surface $C$ for $V\left(\phi\right)=m^{2}\phi^{2}/2$,
in $(a,H,\phi,\dot{\phi})$ coordinates, at slices of various values
of the scale factor $a$. Top row: $\kappa=0.5$, middle row: $\kappa=0$,
bottom row: $\kappa=-0.5$. Units used are $M_{\mathrm{Pl}}=1$ and
$m=0.2$.}
\end{figure*}

\begin{figure*}
\noindent \begin{centering}
\includegraphics[width=1.5in]{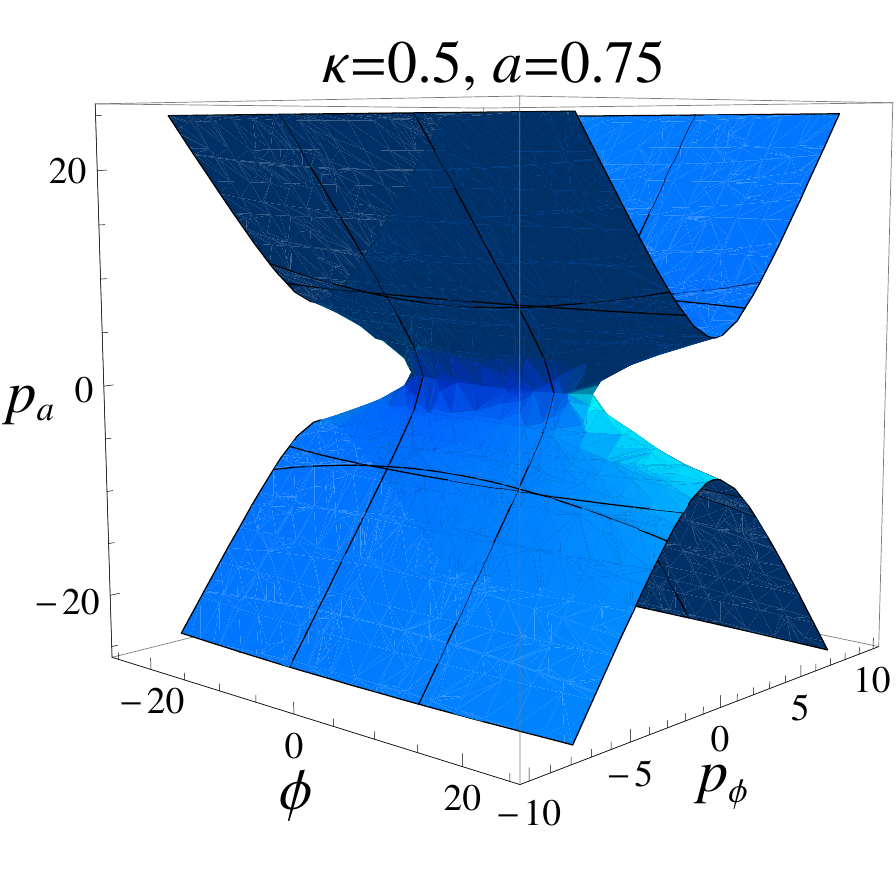}\includegraphics[width=1.5in]{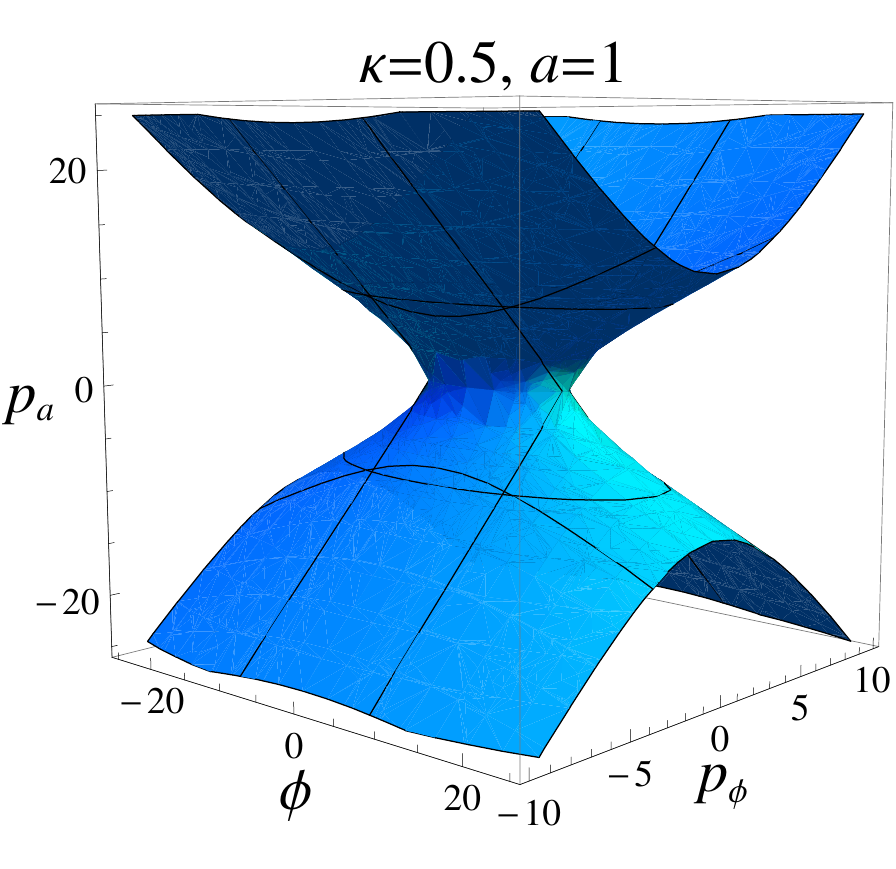}\includegraphics[width=1.5in]{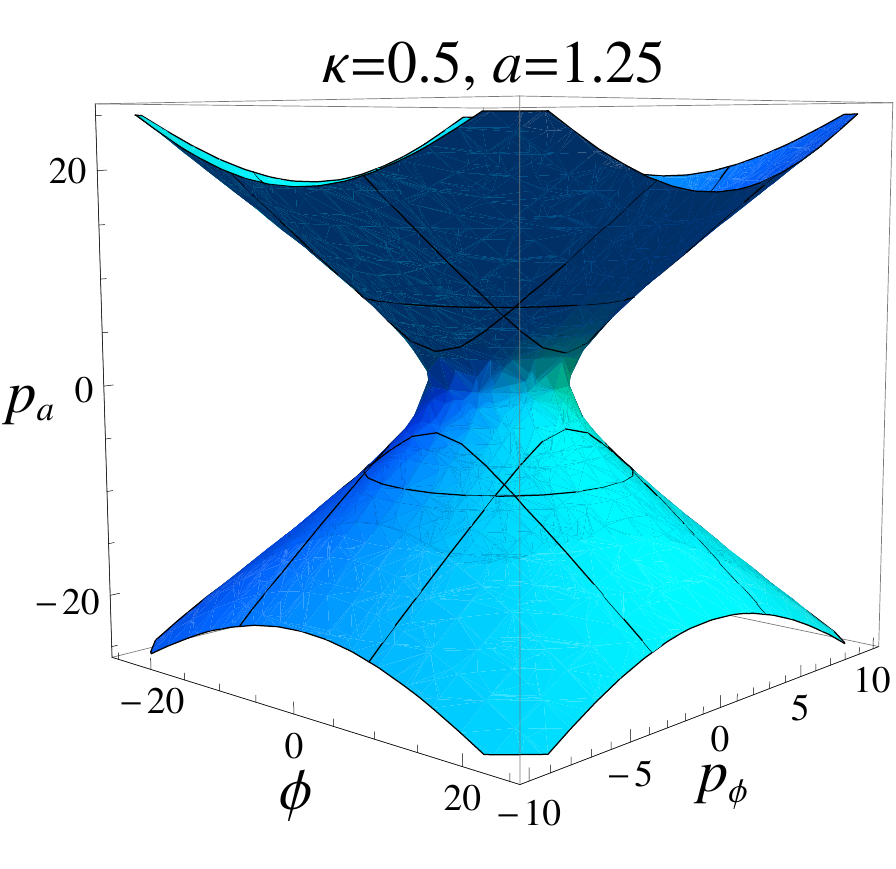}\includegraphics[width=1.5in]{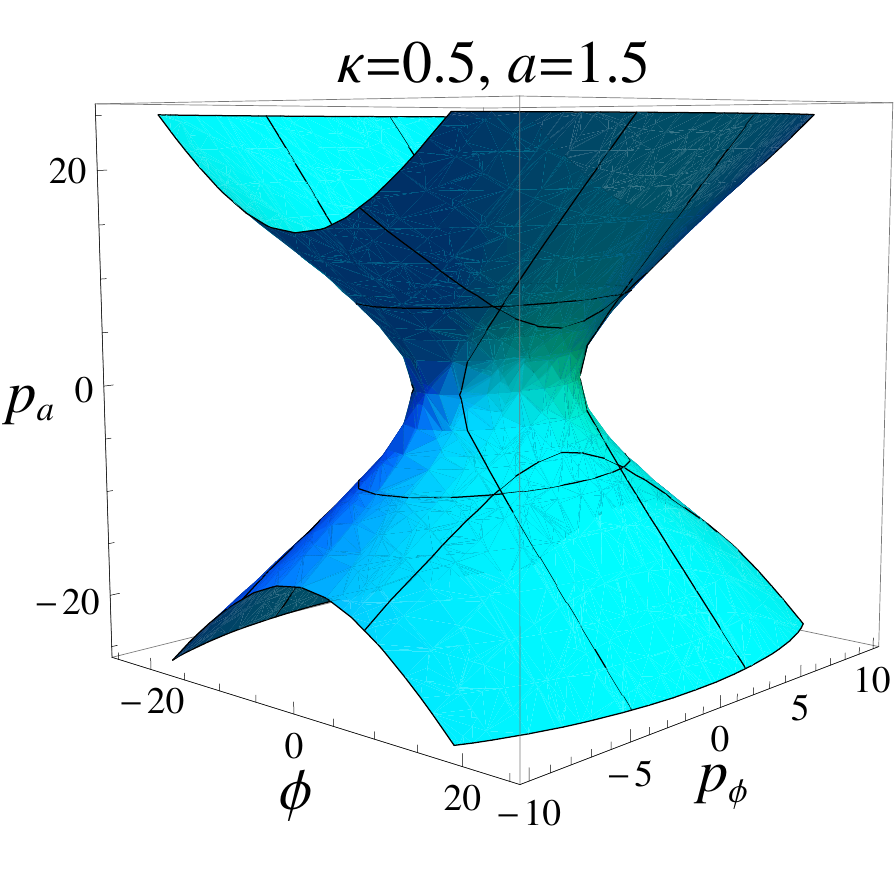} 
\par\end{centering}

\noindent \begin{centering}
\includegraphics[width=1.5in]{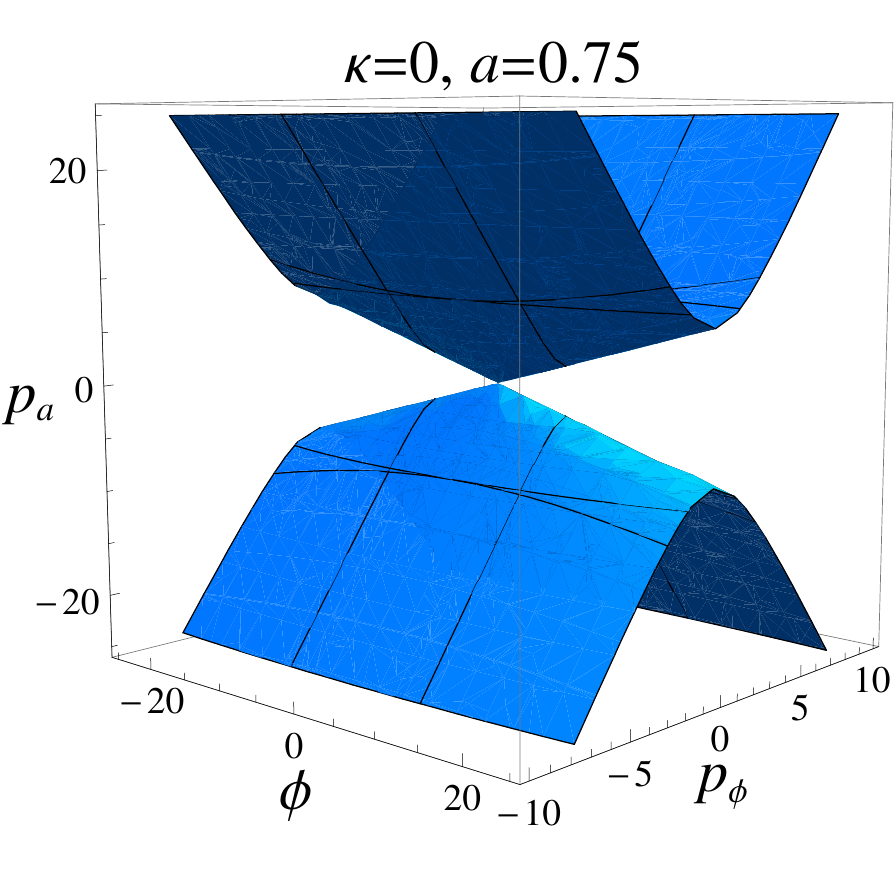}\includegraphics[width=1.5in]{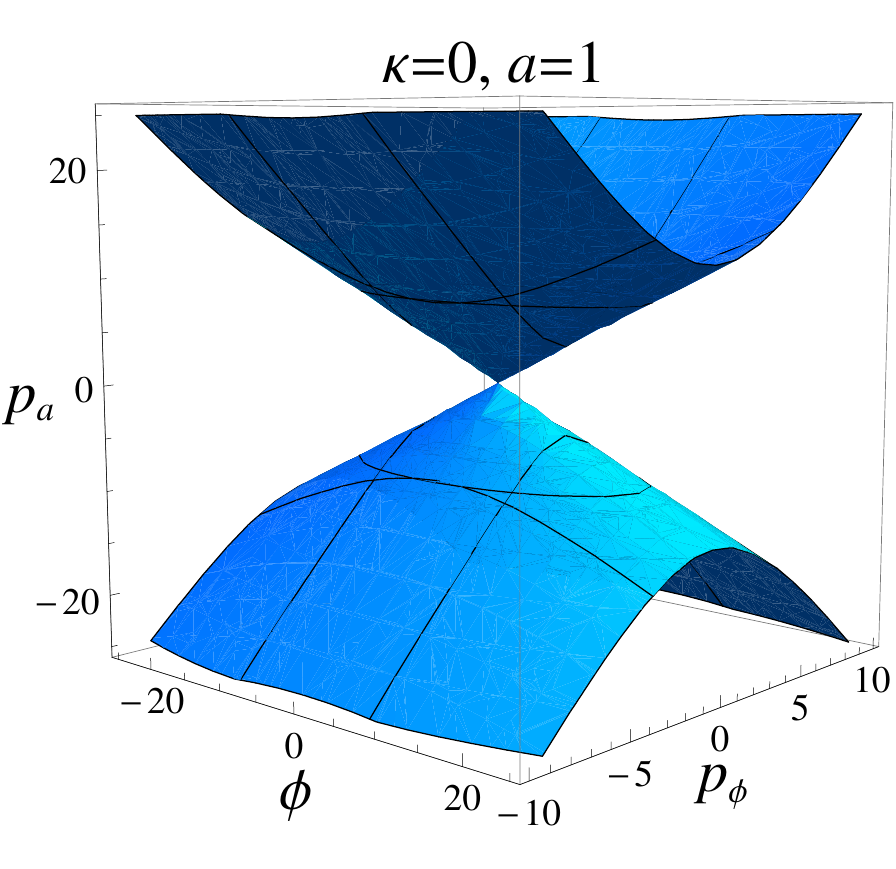}\includegraphics[width=1.5in]{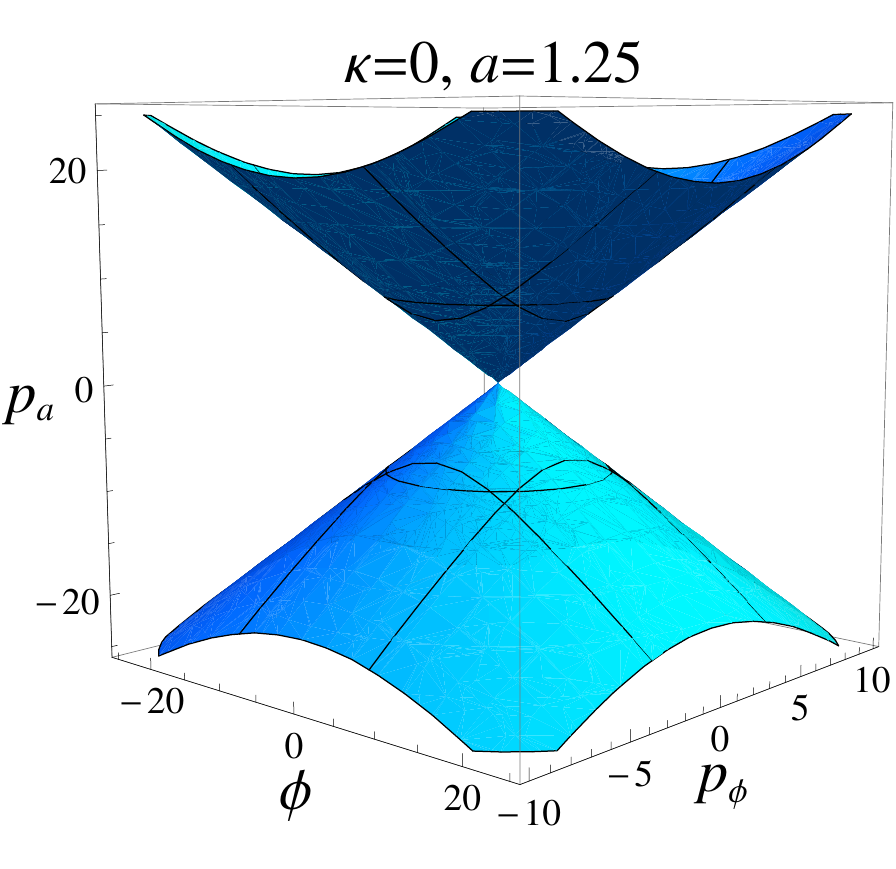}\includegraphics[width=1.5in]{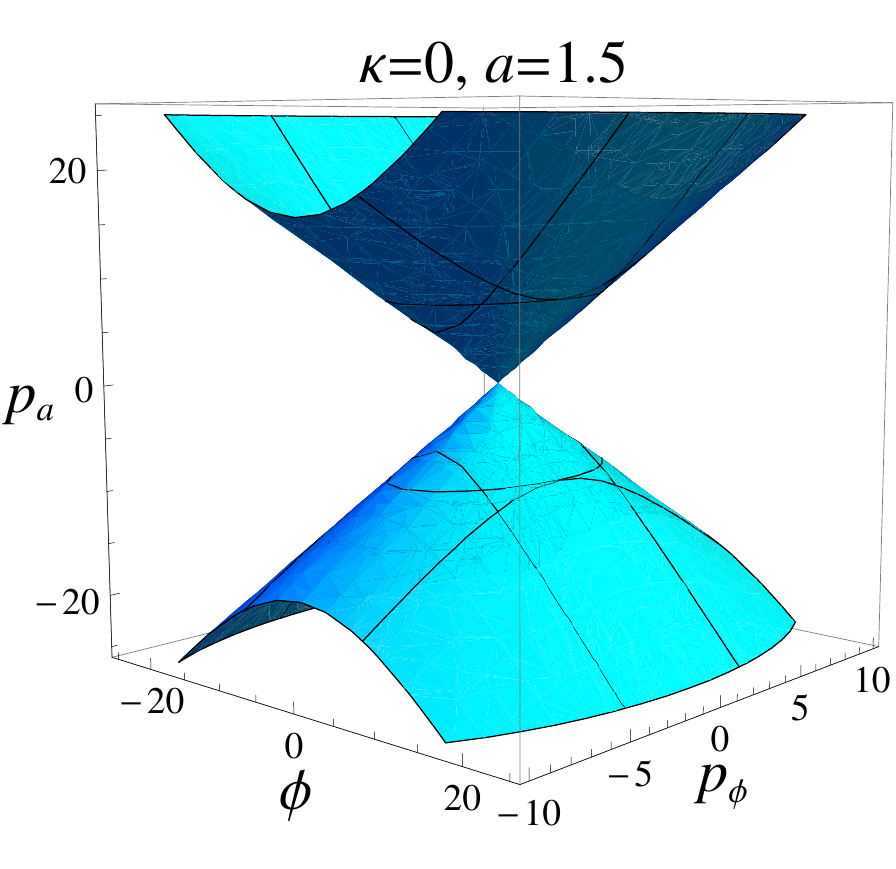} 
\par\end{centering}

\noindent \begin{centering}
\includegraphics[width=1.5in]{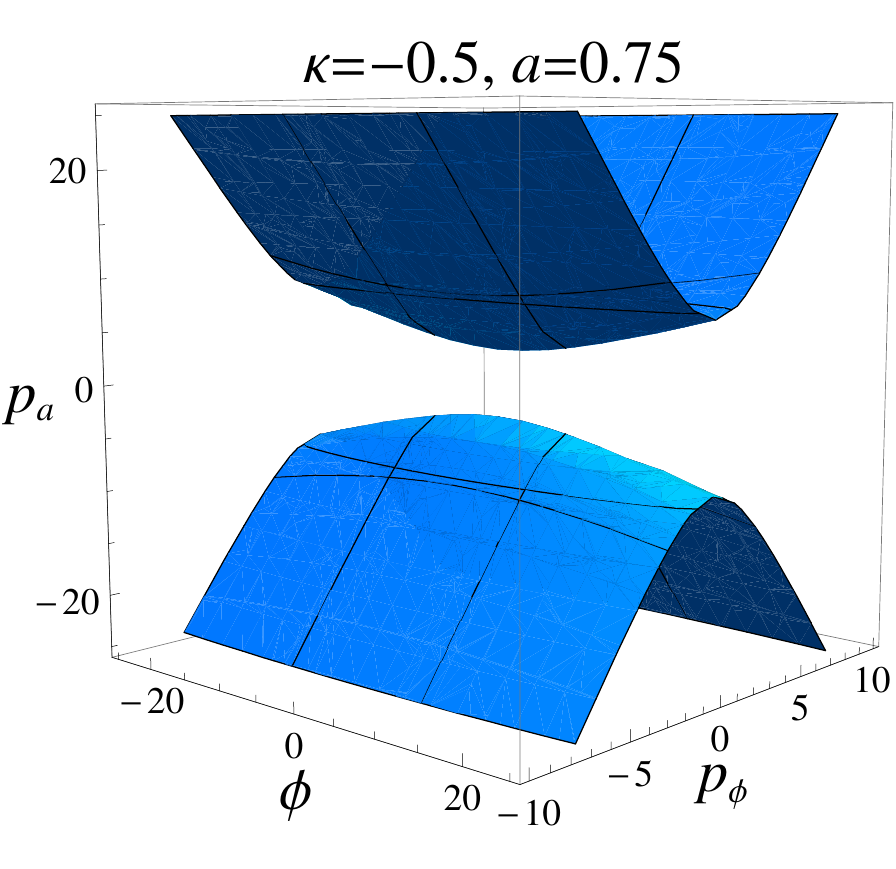}\includegraphics[width=1.5in]{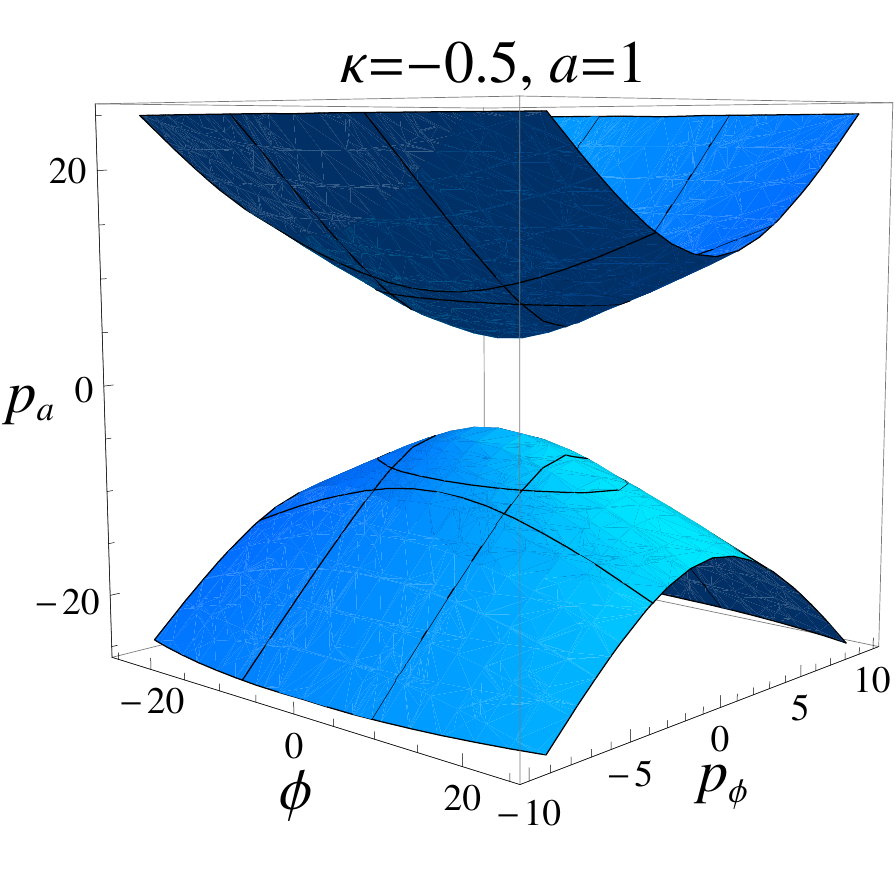}\includegraphics[width=1.5in]{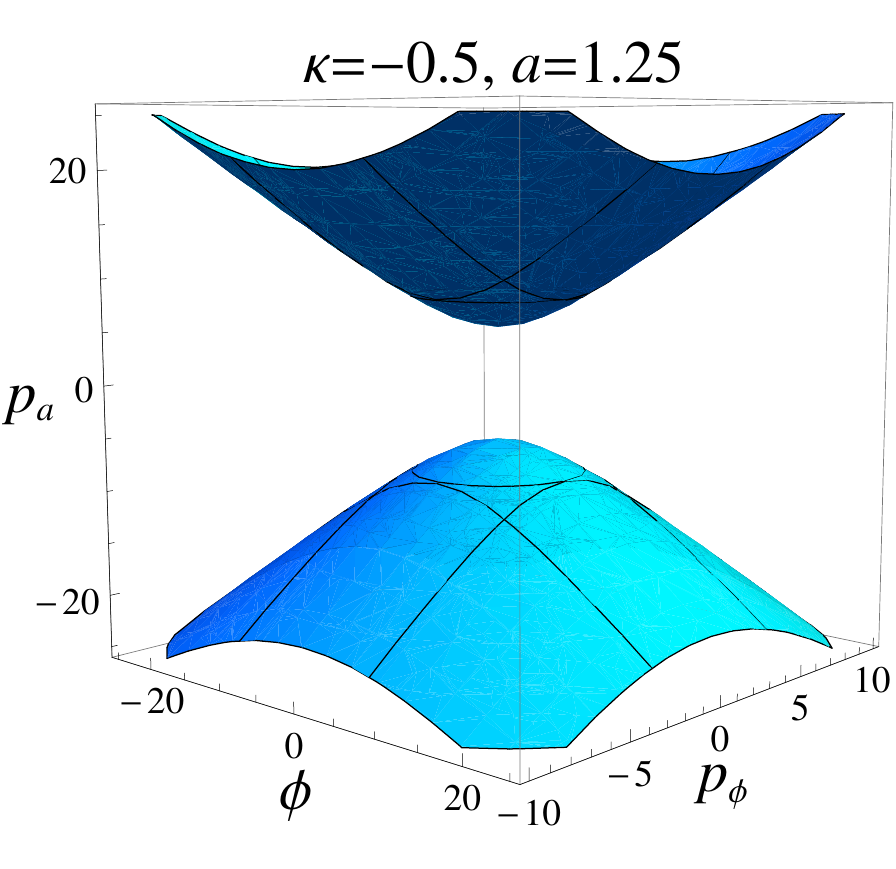}\includegraphics[width=1.5in]{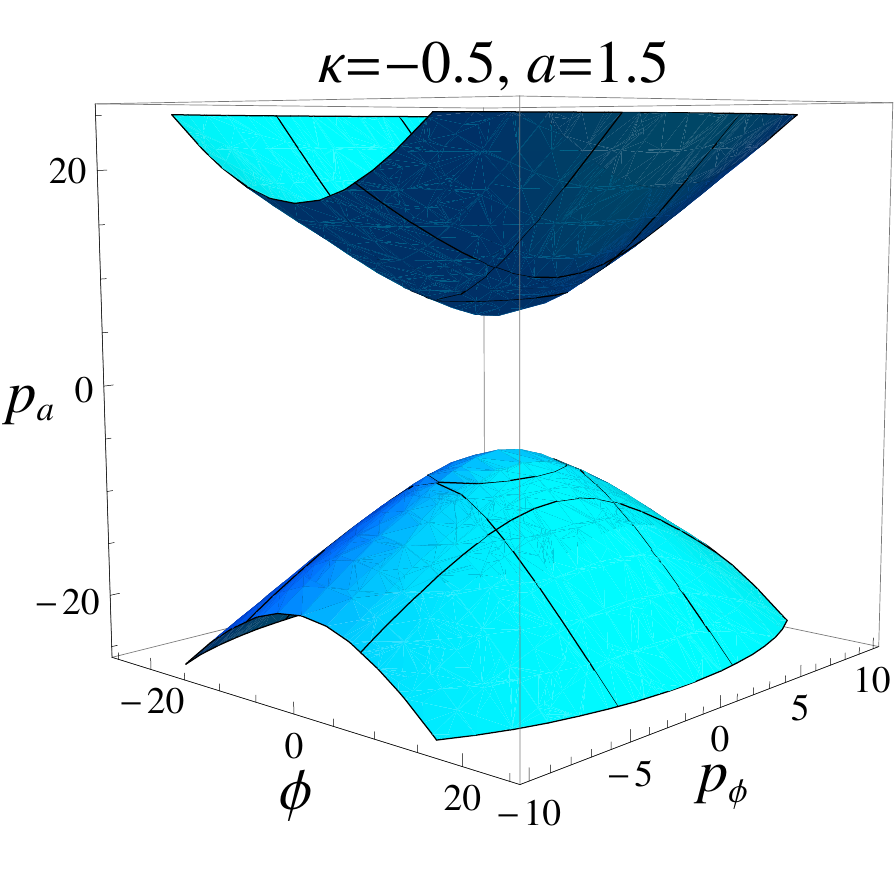} 
\par\end{centering}

\caption{\label{pasurfaces}Plots of the Hamiltonian constraint manifold $C$
for $V\left(\phi\right)=m^{2}\phi^{2}/2$, in $(a,p_{a},\phi,p_{\phi})$
coordinates, in two-dimensional slices $C_{a}$ at various values
of the scale factor $a$. Top row: $\kappa=0.5$, middle row: $\kappa=0$,
bottom row: $\kappa=-0.5$. Units used are $M_{\mathrm{Pl}}=1$ and
$m=0.2$.}
\end{figure*}

Consider a Hamiltonian submanifold $C$ for some choice of curvature
$\kappa$. Restricted to a particular value of the scale factor $a$,
the Hamiltonian submanifold becomes a two-dimensional surface $C_{a}$
immersed within a three-dimensional space $\Gamma_{a}$. We can think
of $C$ as being the disjoint union of the $C_{a}$. Formally speaking,
$C$ is formed by the fibration\emph{ }of the family of manifolds
$C_{a}$ over the positive real line $\mathbb{R}_{+}$ parametrized
by $a$: in general, this produces a non-factorizable three-manifold
within $\Gamma$, since the $C_{a}$ are often different in size and
shape for different values of $a$. As one can see in Fig. \ref{slices},
all the $C_{a}$ are the same if $\kappa=0$. This is due to the fact
that for the choice $\kappa=0$, the Hamiltonian constraint \eqref{FriedmannEquation}
is independent of $a$ in $(a,H,\phi,\dot{\phi})$ coordinates: 
\begin{equation}
H^{2}=\frac{1}{3}\left[\frac{1}{2}\dot{\phi}^{2}+V\left(\phi\right)\right].\label{FriedmannEqflat}
\end{equation}
More precisely, parametrizing $C_{a}=\Gamma_{a}\cap C$ with the other
three coordinates on $\Gamma$ (excluding $a$), we find that, for
$\kappa=0$, $C_{a}$ contains the same set of points, a cone in $\Gamma_{a}$,
regardless of the choice of $a$. Hence, one can pick a two-manifold
$C_{a_{\star}}$ for any choice of scale factor $a_{\star}$ and find
that $C$ is a product space: $C=C_{a_{\star}}\times\mathbb{R}_{+}$.
This is the sense in which $a$ ceases to be a dynamical variable
for flat universes.

As previously, let $K\cong\mathbb{R}^{2}$ denote $\phi$-$\dot{\phi}$
space and consider the vector field invariant map $\chi:\, C\rightarrow K$
defined by $\chi(a,H,\phi,\dot{\phi})=(\phi,\dot{\phi})$, where $C$
is the Hamiltonian submanifold for a flat universe. More generally,
we could let $K$ be any space isomorphic to $\phi$-$\dot{\phi}$
space and let $\chi$ be any function for which the preimage of a
point in $K$ is the set of all points in $C$ with a particular value
of $\phi$ and $\dot{\phi}$. It was previously shown that $\chi$
is vector field invariant with respect to the Hamiltonian flow vector
$X_{\mathcal{H}}$. From Fig. \ref{slices} we can see why this is
true. The Hamiltonian flow vector describes a vector field on $C$;
on each manifold $C_{a}$, the $H$-, $\phi$-, and $\dot{\phi}$-components
of $X_{\mathcal{H}}$ give a vector field, which we can imagine describing
flow tangent to each of the slices shown in Fig. \ref{slices}. The
projection of the vector field from a slice $C_{a}$ down into the
horizontal plane in Fig. \ref{slices} corresponds to the pushforward
of the vector field from $C_{a}$ to $K$. If this vector field in
$K$ is the same no matter which slice $C_{a}$ we chose, then $\chi$
is vector field invariant with respect to $X_{\mathcal{H}}$. This
is manifestly true for the flat universe case, because $C$ factors
as shown above. It is also clear from Fig. \ref{slices} that this
is not true for $\kappa\neq0$: the manifold $C_{a}$ changes dramatically
as $a$ is varied, so the vector field that we push forward to $K$
will be different for different $a$.

At this point we may ask again: is this property of $\phi$ and $\dot{\phi}$
really distinctive? That is, does a different choice of coordinates
on $\Gamma$, say $(a,p_{a},\phi,p_{\phi})$ give the same result:
a map of the form $\psi$ from $(a,p_{a},\phi,p_{\phi})$ to $(\phi,p_{\phi})$
that, provided $\kappa=0$, is vector field invariant with respect
to $X_{\mathcal{H}}$? We saw in Fig. \ref{phipphiplot} that this
is not the case, but it is useful to consider why vector field invariance
fails for $\phi$-$p_{\phi}$ space from the geometrical point of
view. As we see in Fig. \ref{pasurfaces}, even in the $\kappa=0$
case, the partition of the manifold $C$ into $C_{a}$ yields an inequivalent
set of points in $\Gamma_{a}$ for different values of $a$ when parametrized
by $(p_{a},\phi,p_{\phi})$, so that $C$ is merely a fibration of
the $C_{a}$ over $\mathbb{R}_{+}$. Another way of saying this is
that in $(a,p_{a},\phi,p_{\phi})$ coordinates, $C$ is non-factorizable
even in the $\kappa=0$ case. Hence, drawing components of the Hamiltonian
flow vector field on the partition of $C$, we see that a projection
into the $\phi$-$p_{\phi}$ plane will give a vector field that is
different at different values of $a$: that is, $\psi:\,(a,p_{a},\phi,p_{\phi})\rightarrow(\phi,p_{\phi})$
is \emph{not} vector field invariant with respect to $X_{\mathcal{H}}$.
In this way, we have shown that the property of vector field invariance
that $\chi$ possesses is non-trivial and not a generic property of
any map from $C$ onto a two-dimensional manifold: $\phi$ and $\dot{\phi}$
are coordinates with a special property.

\section{Constructing a Measure on Effective Phase Space\label{MeasureConstruction}}

We now have in hand an effective phase space $K$ for flat scalar-field
cosmology, namely, $\phi$-$\dot{\phi}$ space. Its properties are
defined generally through the formalism of a vector field invariant
map and it contains all of the dynamical variables describing the
evolution of a flat FRW universe dominated by a scalar field. However,
while $K$ captures the entire dynamics of the system (every point
is part of a unique trajectory), it is not naturally a symplectic
manifold, the coordinates $(\phi,\dot{\phi})$ are not canonically
conjugate, and there is no reason to expect the na\"{i}ve measure
${\rm d}\dot{\phi}\wedge{\rm d}\phi$ to be conserved. We now ask
whether these features can be corrected, by finding a measure on this
effective phase space that actually is conserved.

While the Liouville measure \eqref{GHSmeasure} is appropriate for
the full phase space $\Gamma$, we are interested now in finding a
measure on the effective phase space. Taking a constructive approach,
we first examine the constraint imposed by conservation of the measure
under Hamiltonian evolution of trajectories, calling such a measure
a ``conserved measure''. We then examine the question of whether
the effective phase space itself has a Lagrangian description, that
is, whether the equation of motion in terms of $\phi$ and $\dot{\phi}$
alone can be derived from a Lagrangian $\mathcal{L}_{K}$ defined
on $K$. If such a Lagrangian exists, it allows us to define a conjugate
momentum $\pi_{\phi}\equiv\partial\mathcal{L}_{K}/\partial\dot{\phi}$.
The measure $\mathrm{d}\pi_{\phi}\wedge\mathrm{d}\phi$ on $K$ is
then automatically conserved under the Hamiltonian flow. We show that
the converse is also true; if there is a conserved measure, there
is a corresponding Lagrangian description. Finally, for the special
case of an $m^{2}\phi^{2}$ potential, we examine the behavior of
the measure at early and late times and prove that the measure on
$K$ exists.

\subsection{Conservation under Hamiltonian flow}

As shown in Ref. \cite{CarrollTam}, the GHS measure \eqref{GHSmeasure}
diverges for flat universes ($\kappa=0$); see also Ref. \cite{GibbonsTurok}.
Specifically, as $\Omega_{k}$, the fraction of the critical energy
density parametrized by curvature, approaches zero, $\Theta\propto\left|\Omega_{k}\right|^{-5/2}$.
In this sense, as \citet{CarrollTam} note, the flatness problem in
cosmology is illusory, a consequence of implicitly assuming a flat
measure on the space of FRW solutions; all universes but a set of
measure zero are spatially flat, according to the GHS measure. This
divergence was briefly noted by \citet{GHS}. The GHS measure is only
well-defined for Hamiltonian systems with an odd number of constraints
(i.e., the Hamiltonian submanifold corresponds to a single constraint).
However, it is important to note that the specification of a flat
universe does not increase the number of constraints, since this just
amounts to selecting a particular value of $\kappa$ in Eq. \eqref{FriedmannEquation}.
See discussion in Sec. \ref{GeometricalPicture} for details of the
phase-space topology.

Given the observed (near-)flatness of our own universe \cite{Planck},
it is well-motivated to consider the question of the measure on the
subspace of $\Gamma$ corresponding to flat universes. Because of
its divergent behavior for $\kappa=0$, the GHS measure cannot help
us in this case. Earlier attempts to regularize the measure, for example
by considering an $\epsilon$-neighborhood around the zero-curvature
Hamiltonian constraint surface \cite{CarrollTam} or by identifying
universes with similar curvatures \cite{GibbonsTurok} have not proven
satisfactory%
\footnote{We thank Alan Guth for conversations on this point.%
}; see also Refs. \cite{SchiffrinWald,HawkingPage}. A different approach
seems to be required. As we have seen, the scale factor $a$ becomes
non-dynamical for $\kappa=0$ and the scalar coordinates $\phi$ and
$\dot{\phi}$ constitute an effective phase space, by virtue of the
vector field invariant mapping discussed in Sec. \ref{EffPhSp}. Though
the GHS and Liouville measures give us no information in this subspace,
we can use the principles and reasoning of the full phase-space argument
to motivate the treatment of the measure question on effective phase
space. As noted in Sec. \ref{MathPrelims}, it is conventional to
implicitly assume a flat measure $\mathrm{d}\dot{\phi}\wedge\mathrm{d}\phi$
in effective phase space when making statements about attractors,
number of $e$-foldings, etc. Considering the measure question in
effective phase space allows us to assess the validity of this assumption.

For simplicity of notation, let the vector field $\tilde{X}_{\mathcal{H}}$
induced from the Hamiltonian evolution vector $X_{\mathcal{H}}$ under
the mapping $\chi:\, C\rightarrow K\cong(\phi,\dot{\phi})$ be written
in $K$ as $\mathbf{v}$. Define $x=\phi$ and $y=\dot{\phi}$. The
measure on a two-dimensional space is a two-form $\boldsymbol{\sigma}$,
which we can always write as 
\begin{equation}
\boldsymbol{\sigma}=f(x,y)\mathrm{d}x\wedge\mathrm{d}y
\end{equation}
for some function $f(x,y)$. We seek a measure that is conserved with
evolution along $\mathbf{v}$: 
\begin{equation}
\pounds_{\mathbf{v}}\boldsymbol{\sigma}=\pounds_{\mathbf{v}}\left[f\left(x,y\right)\mathrm{d}x\wedge\mathrm{d}y\right]=0.\label{Liederiv}
\end{equation}
We can compactly express the condition \eqref{Liederiv} as the vector
equation 
\begin{equation}
\nabla\cdot\left(f\mathbf{v}\right)=0.\label{eulerconst}
\end{equation}
Note that this is equivalent to one of the Euler equations of fluid
dynamics for a steady flow, $\partial\rho/\partial t=0$, where $\rho$
is the density of the fluid and $\mathbf{v}$ its velocity field:
\begin{equation}
\frac{\partial\rho}{\partial t}+\nabla\cdot\left(\rho\mathbf{v}\right)=0.\label{Eulereqn}
\end{equation}
This is simply the statement of conservation of mass. Hence, our conserved
two-form can be thought of as the density of fluid in a steady-flow
system. The probability of a given bundle of trajectories is conserved
under Hamiltonian evolution, just as the mass of a parcel of fluid
is conserved as it flows.

For a single scalar with a canonical kinetic term, the vector field
$\mathbf{v}$ can be found from $X_{\mathcal{H}}$ as follows. Setting
$X_{\mathcal{H}}^{(p_{\phi})}$ given in Eq. \eqref{vectorcomps}
equal to $\partial_{t}p_{\phi}$ {[}recalling from Eq. \eqref{momenta}
that $p_{\phi}=a^{3}\dot{\phi}${]}, we have the Klein--Gordon equation
\begin{equation}
\ddot{\phi}+3H\dot{\phi}+V^{\prime}\left(\phi\right)=0.
\end{equation}
With $H$ as given by the Friedmann equation (the Hamiltonian constraint)
\eqref{FriedmannEqflat}, we have 
\begin{equation}
\ddot{\phi}=-\sqrt{3}\dot{\phi}\sqrt{\frac{1}{2}\dot{\phi}^{2}+V\left(\phi\right)}-V^{\prime}\left(\phi\right)=v^{\dot{\phi}}.\label{eomphi}
\end{equation}
The vector field in $\phi$-$\dot{\phi}$ space is therefore 
\begin{equation}
\mathbf{v}=\left(y,\;-\sqrt{3}y\sqrt{V\left(x\right)+\frac{1}{2}y^{2}}-V^{\prime}\left(x\right)\right).\label{vectorfield}
\end{equation}

\subsection{Existence of a Lagrangian\label{effectiveLagrangian}}

We now have an equation of motion \eqref{eomphi} for $\phi$, obtained
from the Friedmann and Klein--Gordon equations and defined by a potential
$V\left(\phi\right)$. We are looking for a Lagrangian on the effective
phase space $K\cong\left(x,y\right)$ from which an equivalent equation
of motion can be derived. One reason for considering a Lagrangian
description is that the direct approach, i.e., finding a closed-form
solution to the Euler equation \eqref{eulerconst} for the vector
field \eqref{vectorfield}, is highly nontrivial for a typical potential.

The existence of a Lagrangian given an equation of motion is a famous
question known as the inverse problem of the calculus of variations,
which was finally solved by \citet{Douglas} in 1941. See \citet{Santilli}
for further reference. Suppose we have an equation of motion in a
single variable 
\begin{equation}
\ddot{x}=F\left(x,\dot{x}\right).\label{eomset}
\end{equation}
Then Douglas' theorem states that there exists a Lagrangian for which
the Euler--Lagrange equation gives the correct equation of motion
\eqref{eomset} if and only if there exists a function $f$ satisfying
the Helmholtz condition 
\begin{equation}
\frac{\mathrm{d}f}{\mathrm{d}t}+\frac{\partial F}{\partial\dot{x}}f=0,\label{Helmholtz}
\end{equation}
or equivalently, with $y=\dot{x}$, 
\begin{equation}
\frac{\partial f}{\partial t}+\frac{\partial}{\partial x}\left(\dot{x}f\right)+\frac{\partial}{\partial y}\left(Ff\right)=0.
\end{equation}

For the problem at hand, defining $\phi=x$ and $\dot{\phi}=y$ as
before, Eq. \eqref{eomphi} can be written 
\begin{equation}
F\left(x,y\right)=-\sqrt{3}y\sqrt{\frac{1}{2}y^{2}+V\left(x\right)}-V^{\prime}\left(x\right)=\ddot{x}.\label{eomphiprime}
\end{equation}
Noting from Eq. \eqref{vectorfield} that $\mathbf{v}=\left(y,\; F\right)$,
we are able to write the Helmholtz condition in the form 
\begin{equation}
\frac{\partial f}{\partial t}+\frac{\partial}{\partial x}\left(v^{x}f\right)+\frac{\partial}{\partial y}\left(v^{y}f\right)=\frac{\partial f}{\partial t}+\nabla\cdot\left(f\mathbf{v}\right)=0.\label{Euler2}
\end{equation}
This is precisely the Euler equation for fluid flow \eqref{Eulereqn},
with $f$ taking the place of the density. If there is a measure $f\;\mathrm{d\dot{\phi}}\wedge\mathrm{d}\phi$
on $\phi$-$\dot{\phi}$ space conserved along the Hamiltonian flow
vector, then $\nabla\cdot\left(f\mathbf{v}\right)=0$. Thus, we have
proven the following: 
\begin{quote}
There exists a Hamiltonian-flow conserved measure on $\phi$-$\dot{\phi}$
space if and only if the equation of motion $\ddot{\phi}+\sqrt{3}\dot{\phi}\sqrt{\dot{\phi}^{2}/2+V\left(\phi\right)}+V^{\prime}\left(\phi\right)=0$
possesses a Lagrangian description in effective phase space. More
specifically, there exists a time-independent measure on $\phi$-$\dot{\phi}$
space if and only if the Helmholtz condition is satisfied by a time-independent
function $f(\phi,\dot{\phi})$. 
\end{quote}
In other words, a Lagrangian description of the equation of motion
\eqref{eomphiprime} {[}cf. Eq. \eqref{eomphi}{]} exists if and only
if the Helmholtz condition \eqref{Helmholtz} is satisfied. In turn,
the Helmholtz condition is satisfied if and only if there is a function
$f$ satisfying the Euler equation \eqref{Eulereqn} for fluid flow
with the Hamiltonian vector field \eqref{vectorfield}. Whether or
not there is such a function is difficult to establish in general,
although we will give an argument in the case of $m^{2}\phi^{2}$
potentials.

\subsection{The conjugate momentum and the measure\label{momentum&measure}}

If the Helmholtz condition \eqref{Helmholtz} is satisfied, then the
equation of motion can be written in the form 
\begin{equation}
A\left(t,\phi,\dot{\phi}\right)\ddot{\phi}+B\left(t,\phi,\dot{\phi}\right)=0,\label{eomsantilli}
\end{equation}
where 
\begin{equation}
\frac{\partial B}{\partial\dot{\phi}}=\left(\frac{\partial}{\partial t}+\dot{\phi}\frac{\partial}{\partial\phi}\right)A.\label{self-adjointness}
\end{equation}
Explicitly, given $f$ satisfying Eq. \eqref{Euler2} and with $A=f(t,\phi,\dot{\phi})$
and $B=-f(t,\phi,\dot{\phi})\, F(\phi,\dot{\phi})$, where $F$ is
given in Eq. \eqref{eomphiprime}, it can be shown that Eq. \eqref{self-adjointness}
is satisfied and that Eq. \eqref{eomsantilli} corresponds to the
correct equation of motion \eqref{eomphi}. Then as shown in Ref.
\cite{Santilli}, the Lagrangian can be constructed explicitly: 
\begin{equation}
\mathcal{L}_{K}\left(t,\phi,\dot{\phi}\right)=G\left(t,\phi,\dot{\phi}\right)+C\left(t,\phi\right),\label{Lagr}
\end{equation}
where 
\begin{equation}
\begin{aligned}G\!\left(t,\!\phi,\!\dot{\phi}\right)\! & =\!\dot{\phi}\!\int_{0}^{1}\!\!\mathrm{d}\tau^{\prime}\!\!\left[\dot{\phi}\!\int_{0}^{1}\!\!\mathrm{d}\tau A\!\!\left(t,\!\phi,\!\tau\dot{\phi}\right)\right]\!\!\left(t,\!\phi,\!\tau^{\prime}\dot{\phi}\right),\\
C\!\left(t,\!\phi\right)\! & =\!\phi\int_{0}^{1}\mathrm{d}\tau\, W\left(t,\!\tau\phi\right),\;\;\;\mathrm{and}\\
W\!\left(t,\!\phi\right)\! & =\!-B-\frac{\partial G}{\partial\phi}+\frac{\partial^{2}G}{\partial\dot{\phi}\partial t}+\frac{\partial G}{\partial\phi\partial\dot{\phi}}\dot{\phi}.
\end{aligned}
\label{lagrSantilli}
\end{equation}

We can then extract the momentum $\pi_{\phi}$ conjugate to $\phi$
in $\phi$-$\dot{\phi}$ space via 
\begin{equation}
\pi_{\phi}=\frac{\partial\mathcal{L}_{K}}{\partial\dot{\phi}}=\frac{\partial G}{\partial\dot{\phi}}.\label{piphi}
\end{equation}
Then the Liouville measure on $\phi$-$\dot{\phi}$ space is 
\begin{equation}
\mathrm{d}\pi_{\phi}\wedge\mathrm{d}\phi=\frac{\partial^{2}G}{\partial\dot{\phi}^{2}}\mathrm{d}\dot{\phi}\wedge\mathrm{d}\phi.
\end{equation}
With $A=f$, it can be shown from Eq. \eqref{lagrSantilli} that $\partial_{\dot{\phi}}^{2}G=f$
and hence 
\begin{equation}
\mathrm{d}\pi_{\phi}\wedge\mathrm{d}\phi=f\;\mathrm{d}\dot{\phi}\wedge\mathrm{d}\phi.
\end{equation}

In the previous section, we demonstrated that finding a Lagrangian
producing the equation of motion on $\phi$-$\dot{\phi}$ space was
the same problem as constructing a Hamiltonian-flow conserved measure
on that space. The result we have proven in this section states something
stronger: 
\begin{quote}
The natural Liouville measure $\mathrm{d}\pi_{\phi}\wedge\mathrm{d}\phi$
on effective phase space that one obtains from the effective Lagrangian,
if it exists, is the same measure that one finds by explicitly constructing
a nontrivial two-form $f\,\mathrm{d}\dot{\phi}\wedge\mathrm{d}\phi$
conserved under Hamiltonian evolution. 
\end{quote}
We note that these results are applicable to any single-field $V\left(\phi\right)$
model with canonical kinetic term, or with slight generalization,
to any dynamical problem in a single variable.

\section{The Measure on Effective Phase Space for Quadratic Potentials}

It is illustrative to explicitly investigate the behavior of the measure
on effective phase space for a specific model, 
\begin{equation}
V\left(\phi\right)=\frac{1}{2}m^{2}\phi^{2}.
\end{equation}
Ideally, one would like to obtain a closed-form expression for the
measure; however, solving the partial differential equation explicitly
proves to be prohibitive. We obtain expressions for the behavior of
the measure in two limits, 
$H\gg m$ and $H\ll m$, which can be viewed as corresponding to early
and late times. Finally, we use the Cauchy--Kowalevski theorem to
prove that a unique measure obeying the constraint \eqref{eulerconst}
exists, up to overall normalization.

\subsection{Constraining the behavior of the measure}

It is convenient at this point to reparametrize the vector field in
terms of polar coordinates. Again setting $x=\phi$ and $y=\dot{\phi}$,
let 
\begin{equation}
r=\sqrt{y^{2}+m^{2}x^{2}}=\sqrt{6}H
\end{equation}
and 
\begin{equation}
\tan\theta=\frac{y}{mx}.
\end{equation}
Then the vector field \eqref{vectorfield} can be written as 
\begin{equation}
\mathbf{v}=-\sqrt{\frac{3}{2}}r^{2}\sin^{2}\theta\;\boldsymbol{\hat{\mathbf{r}}}-\left(mr+\sqrt{\frac{3}{2}}r^{2}\sin\theta\cos\theta\right)\boldsymbol{\hat{\theta}},\label{vpolar}
\end{equation}
where $\mathbf{\hat{r}}=\mathbf{\hat{x}}m^{-1}\cos\theta+\mathbf{\hat{y}}\sin\theta$
and $\boldsymbol{\hat{\theta}}=-\mathbf{\hat{x}}m^{-1}\sin\theta+\mathbf{\hat{y}}\cos\theta$
are unit vectors under the appropriate scaling of axes. The constraint
\eqref{eulerconst} for a time-independent measure $f$ may be expressed
as an explicit partial differential equation 
\begin{equation}
\begin{aligned}0 & =-\frac{1}{r}\sqrt{\frac{3}{2}}\sin^{2}\theta\partial_{r}\left(r^{3}f\right)\\
 & \hspace{1em}\hspace{1em}-m\partial_{\theta}f-\sqrt{\frac{3}{2}}r\partial_{\theta}\left(f\sin\theta\cos\theta\right).
\end{aligned}
\label{explicitPDE}
\end{equation}

\subsubsection{Late universe limit}

The late-universe, $H\rightarrow0$ limit corresponds to $r\rightarrow0$.
Suppose first that $f$ does not diverge in this limit. Then as $\mathbf{v}\cdot\mathbf{\hat{r}}=-\sqrt{3/2}r^{2}\sin^{2}\theta<0$
for all $r$, $\theta$, it follows that for any circular disk $R$
of radius $r_{\star}$ centered at the origin, 
\begin{equation}
\begin{aligned}\int_{R}\!\nabla\!\cdot\!\left(f\mathbf{v}\right)\!\mathrm{d}A & =\oint_{\partial R}\!\!\left(f\mathbf{v}\right)\!\cdot\!\mathbf{\hat{n}}\mathrm{d}\ell\\
 & =-\sqrt{\frac{3}{2}}\int_{0}^{2\pi}\!\! f\left(r_{\star},\theta\right)r_{\star}^{3}\sin^{2}\theta\,\mathrm{d}\theta.
\end{aligned}
\end{equation}
Since the expression on the right-hand side is negative, $\nabla\cdot\left(f\mathbf{v}\right)$
is not identically zero. If, however, $f$ diverges as $r\rightarrow0$,
we must include another boundary term: effectively, the disk becomes
an annulus, with the point at $r=0$ removed. In general, this does
not allow us to show that $\nabla\cdot\left(f\mathbf{v}\right)\not\equiv0$.
Thus, we conclude that $f$ must diverge as $r\rightarrow0$.

Near the origin, where $r$ is small, physically corresponds to small
Hubble parameter in units of the scalar field mass, $H\ll m$. 
In this limit, we may take the leading order in $r$ in Eq. \eqref{explicitPDE},
as this will dwarf all other terms: 
\begin{equation}
\partial_{\theta}\left(mf\right)\xrightarrow[r\rightarrow0]{}0.\label{nearoriginindeptheta}
\end{equation}
This means that $f$ is well-behaved in its angular coordinate near
the origin: we do not have any ambiguity in defining $f$ as $r\rightarrow0$
as would occur for, e.g., $f\rightarrow\sin\theta$. Near the origin,
we have $f\rightarrow p\left(r\right)$, where $p\left(r\right)$
is the leading (i.e., most divergent) part of $f$. In general, 
\begin{equation}
f\left(r,\theta\right)=p\left(r\right)+q\left(r,\theta\right),
\end{equation}
where $q$ is subleading as $r\rightarrow0$.

Thus, Eq. \eqref{explicitPDE} implies that for small $r$, 
\begin{equation}
\begin{aligned}0 & =\sin^{2}\theta\left(r^{3}p^{\prime}+3r^{2}p\right)+\sqrt{\frac{2}{3}}mr\partial_{\theta}q\\
 & \hspace{1em}\hspace{1em}+r^{2}p\left(\cos^{2}\theta-\sin^{2}\theta\right),
\end{aligned}
\label{highorder}
\end{equation}
as all other terms, e.g., $\partial_{r}(r^{3}q)$, are of lesser order
in $r$. A solution also has the requirement that $q\left(r,\theta\right)$
be periodic in $\theta$. We obtain a solution to Eq. \eqref{highorder}
if 
\begin{equation}
r^{3}p^{\prime}+3r^{2}p=0
\end{equation}
and 
\begin{equation}
\sqrt{\frac{2}{3}}mr\partial_{\theta}q+r^{2}p\left(\cos^{2}\theta-\sin^{2}\theta\right)=0,
\end{equation}
which imply 
\begin{equation}
p=\frac{C}{r^{3}}\label{solution1}
\end{equation}
and 
\begin{equation}
q=-\sqrt{\frac{3}{2}}\frac{C}{mr^{2}}\sin\theta\cos\theta.\label{solution1'}
\end{equation}
Thus, there exists a solution to Eq. \eqref{explicitPDE} such that
as $r\rightarrow0$, 
\begin{equation}
f\rightarrow C\left(\frac{1}{r^{3}}-\sqrt{\frac{3}{2}}\frac{\sin\theta\cos\theta}{mr^{2}}\right).\label{lateU}
\end{equation}
As the small $r$ form \eqref{lateU} of the measure is divergent,
with degree greater than $2$, it is not normalizable over a region
containing the point $\phi=\mbox{\ensuremath{\dot{\phi}}}=0$. However,
as we shall see in Sec. \ref{existence}, excising the origin and
considering the measure on the punctured plane allows us to prove
that the measure exists and is well-defined for $(\phi,\dot{\phi})\neq(0,0)$.

A question that remains is whether this solution is unique, i.e.,
whether a nontrivial solution to Eq. \eqref{explicitPDE} must have
the behavior \eqref{lateU} near the origin. Writing any near-origin
solution as $f\left(r,\theta\right)=p\left(r\right)+q\left(r,\theta\right)$
as above and demanding that $q\left(r,\theta\right)$ be periodic
in $\theta$ implies that $\partial_{\theta}q$ must also be periodic
and have zero mean, i.e., $T^{-1}\int_{0}^{T}(\partial_{\theta}q)\,\mathrm{d}\theta\xrightarrow[T\rightarrow\infty]{}0$.
But from Eq. \eqref{highorder} we have

\begin{equation}
\begin{aligned}\partial_{\theta}q= & -\sqrt{\frac{3}{2}}\frac{1}{m}\Bigl[rp\left(\cos^{2}\theta-\sin^{2}\theta\right)\\
 & \hspace{1em}\hspace{1em}\hspace{1em}\hspace{1em}\hspace{1em}+\sin^{2}\theta\left(r^{2}p^{\prime}+3rp\right)\Bigr].
\end{aligned}
\end{equation}
At fixed $r$, this expression has zero mean only if $r^{2}p^{\prime}+3rp=0$.
Hence, the solution in Eqs. \eqref{solution1} and \eqref{solution1'}
is unique. That is, any nontrivial solution to Eq. \eqref{explicitPDE}
must have the form \eqref{lateU} as $r\rightarrow0$.

\subsubsection{Early universe limit}

In the large $r$ limit, which corresponds to 
 $H\gg m$, we take the vector field \eqref{vpolar} at leading order
in $r$: 
\begin{equation}
\mathbf{v}\approx-\sqrt{\frac{3}{2}}r^{2}\sin^{2}\theta\;\boldsymbol{\hat{\mathbf{r}}}-\sqrt{\frac{3}{2}}r^{2}\sin\theta\cos\theta\;\boldsymbol{\hat{\theta}}.
\end{equation}
Hence, the Euler constraint \eqref{eulerconst} {[}explicitly, Eq.
\eqref{explicitPDE}{]} requires, for large $r$, that the measure
satisfy 
\begin{equation}
\partial_{\theta}f=-r\tan\theta\partial_{r}f-\left(2\tan\theta+\cot\theta\right)f.
\end{equation}
For $f$ to be periodic in $\theta$ with period $2\pi$ for fixed
$r$, we must have $\partial_{\theta}f$ periodic in $\theta$ with
zero mean (i.e., $\partial_{\theta}f$ oscillates about zero). This
requirement is satisfied by the odd functions $\tan\theta$ and $2\tan\theta+\cot\theta$,
so $\partial_{r}f$ must be periodic in $\theta$ with period $2\pi$.
We therefore take $f\left(r,\theta\right)$ separable as an ansatz:
\begin{equation}
f\left(r,\theta\right)=R\left(r\right)\Theta\left(\theta\right).
\end{equation}
Hence, 
\begin{equation}
0=r\frac{\partial_{r}R}{R}+3+\frac{\partial_{\theta}\left(\Theta\sin\theta\cos\theta\right)}{\Theta\sin^{2}\theta}
\end{equation}
which has solutions 
\begin{equation}
R=Cr^{\gamma-3}
\end{equation}
and 
\begin{equation}
\Theta=C\csc\theta\cos^{\gamma-1}\theta,
\end{equation}
where $\gamma\in\mathbb{R}$ is arbitrary. Demanding that $f$ be
positive everywhere, we have the leading order behavior 
\begin{equation}
f\rightarrow Cr^{\gamma-3}\left|\frac{\cos^{\gamma-1}\theta}{\sin\theta}\right|\;\;\;\mathrm{as}\;\;\; r\rightarrow\infty.\label{earlyU}
\end{equation}
The large $r$ behavior for finite mass $m$ is a weighted sum or
integral of this family of solutions, determined by matching onto
the measure for intermediate values of $r$. The coefficients of the
sum must be found numerically for a particular value of $m$. Note
that the $r\rightarrow\infty$ behavior of $f$ given in Eq. \eqref{earlyU}
diverges near $y\rightarrow0$. This corresponds to the apparent attractor
solution plotted in Fig. \ref{attractors}: on large scales in effective
phase space, the vector field points toward the $\phi$ axis, toward
the apparent attractors near $\dot{\phi}=\pm\sqrt{2/3}m$. Any trajectory
that starts out at large $r$ is driven toward one of these attractor
curves, which on large scales (equivalently, for small $m$) are very
near $\dot{\phi}=0$. Hence, the behavior of this solution is physically
sensible. Imposing the condition $\gamma<1$ makes the radial integral
over large $r$ convergent; this restriction could be viewed as physically
reasonable, as evolving any universe backward must result in $H\rightarrow\infty$,
i.e.,\emph{ }the Big Bang, in finite time.

\subsubsection{The measure near the apparent attractor}

Comparing the $r\rightarrow0$ behavior \eqref{lateU} and the $r\rightarrow\infty$
behavior \eqref{earlyU} of the measure, we see that in both limits,
$f$ diverges wherever trajectories converge in $\phi$-$\dot{\phi}$
space: for the early universe (large $r$) this occurs along the apparent
attractor solution, approximated by $\dot{\phi}\approx0$, while for
small $r$ this occurs at the origin, which corresponds to reheating.
Therefore, it is well-motivated to suppose that any solution to the
full constraint equation $\nabla\cdot\left(f\mathbf{v}\right)=0$
diverges along the full apparent attractor (Fig. \ref{attractors}),
i.e., the curves in $\phi$-$\dot{\phi}$ space satisfying 
\begin{equation}
\frac{\mathrm{d}\dot{\phi}}{\mathrm{d}\phi}=-\frac{\sqrt{\frac{3}{2}}\sqrt{m^{2}\phi^{2}+\dot{\phi}^{2}}\dot{\phi}+m^{2}\phi}{\dot{\phi}}
\end{equation}
with initial condition $\dot{\phi}=\pm\sqrt{2/3}m$, $\phi\rightarrow\mp\infty$.
It is interesting to note the following similarity between the expressions
\eqref{earlyU} and \eqref{lateU} for the measure in the early and
late universe. Defining $d$ as the distance to the apparent attractor
in the $\phi$-$\dot{\phi}$ plane and considering successive ringdown
orbits, it is possible to show that $f\sim1/rd$ during reheating.
Similarly, the $\gamma=1$ solution to Eq. \eqref{earlyU} also corresponds
to $1/rd$ in the early universe. It is not clear and appears unlikely
that these attractor solutions can be written in analytical form,
which would imply that there is no analytical expression for the Lagrangian
or measure.

\subsection{Existence of the measure for $m^{2}\phi^{2}$ potentials\label{existence}}

A natural question to ask, given the constraint imposed on a time-independent
measure on the effective phase space $K$, is whether a non-trivial
solution exists, i.e., does there exist a probability distribution
$f$ satisfying $\nabla\cdot\left(f\mathbf{v}\right)=0$ for $\mathbf{v}$
given in Eq. \eqref{vectorfield}? In general, the answer to this
question is dependent on the potential $V\left(\phi\right)$, but
in light of the results \eqref{lateU} and \eqref{earlyU}, we will
see that we can answer the question in the affirmative for an $m^{2}\phi^{2}$
potential.

As in Eq. \eqref{explicitPDE}, we express the partial differential
equation that $f$ must satisfy in polar coordinates in effective
phase space. Define a function 
\begin{equation}
g\left(r,\theta\right)=r^{3}\sin^{2}\theta\left[f\left(r,\theta\right)-f_{\epsilon}\right],
\end{equation}
where $f_{\epsilon}\equiv f\left(r\rightarrow\epsilon\right)$ is
a constant, for some small $\epsilon>0$. This expression is well-defined
because, as shown in Eq. \eqref{nearoriginindeptheta}, we must have
$f\left(r,\theta\right)\xrightarrow[r\rightarrow0]{}f\left(r\right)$,
independent of $\theta$. We are thinking of the solution for $g$
as a Cauchy problem, with initial data 
\begin{equation}
g\left(r=\epsilon,\theta\right)=0\label{Cauchydata}
\end{equation}
and evolution in $r$ rather than $t$.

The constraint equation \eqref{explicitPDE} for $f$ becomes, in
terms of $g$, 
\begin{equation}
\begin{aligned}\partial_{r}g & =\sin^{2}\theta\partial_{r}\left(r^{3}f\right)\\
 & =-\sqrt{\frac{2}{3}}mr\partial_{\theta}f-r^{2}\partial_{\theta}\left(f\sin\theta\cos\theta\right)\\
 & =-\sqrt{\frac{2}{3}}\frac{m}{r^{2}}\partial_{\theta}\left(g\csc^{2}\theta\right)-\frac{1}{r}\partial_{\theta}\left(g\cot\theta\right).
\end{aligned}
\end{equation}
Defining 
\begin{equation}
a\left(r,\theta\right)=-\sqrt{\frac{2}{3}}\frac{m}{r^{2}}\csc^{2}\theta-\frac{1}{r}\cot\theta
\end{equation}
and 
\begin{equation}
b\left(\theta,g\right)=\left(\frac{1}{r}+2\sqrt{\frac{2}{3}}\frac{m}{r^{2}}\cot\theta\right)g\csc^{2}\theta,
\end{equation}
the constraint becomes 
\begin{equation}
\partial_{r}g=a\left(r,\theta\right)\partial_{\theta}g+b\left(\theta,g\right).\label{Cauchyeqn}
\end{equation}
A measure on $\phi$-$\dot{\phi}$ space exists if and only if there
is a solution $g$ to the evolution equation \eqref{Cauchyeqn} with
initial data \eqref{Cauchydata}.

The function $a$ is analytic in the entire upper half-plane $\mathbb{R}_{+}^{2}$
($0<\theta<\pi$, corresponding to $\dot{\phi}>0$). The function
$b$ is analytic in terms of $r$ in $\mathbb{R}_{+}^{2}$ and is
analytic in terms of $g$ near $g=0$. Since the upper half-plane
is open, the Cauchy--Kowalevski theorem \cite{Kowalevski} guarantees
that there exists a unique analytic solution to the evolution equation
\eqref{Cauchyeqn} in $\mathbb{R}_{+}^{2}$. We note that the specification
of the initial data \eqref{Cauchydata}, in setting the constant $f_{\epsilon}$,
amounts to simply selecting the constant $C$ such that $f\rightarrow Cr^{-3}$
{[}cf. Eq. \eqref{lateU}{]} near the origin. This argument also holds
separately for the lower half-plane $\mathbb{R}_{-}^{2}$ ($\pi<\theta<2\pi$,
corresponding to $\dot{\phi}<0$). The divergence in $a$ and $b$
at $\dot{\phi}=0$ is a coordinate singularity if $\phi\neq0$; the
vector field is finite and trajectories are smooth as they cross the
$\phi$ axis. We can impose continuity of the measure to select the
same normalization in the upper and lower half-planes. The Cauchy--Kowalevski
theorem thus guarantees existence and uniqueness everywhere for $r>\epsilon$.
Strictly speaking, for any particular, finite $\epsilon$, there will
be higher-order corrections to the form $Cr^{-3}$ of the measure
(i.e., terms less divergent than $r^{-3}$), as computed in Eq. \eqref{lateU}.
However, the overall normalization $C$ remains well-defined, since
we can also use the Cauchy--Kowalevski theorem to guarantee uniqueness
upon evolving the measure inward for $r<\epsilon$. We can now consider
a family of such Cauchy problems, for different values of $\epsilon$
but all with the same value of $C$. This family of measures is uniformly
convergent as $\epsilon\rightarrow0$, converging to the form $Cr^{-3}$
near the origin. Thus, the Cauchy--Kowalevski theorem guarantees the
existence, and uniqueness up to normalization, of the measure on the
entire punctured plane $\mathbb{R}^{2}\backslash\{(\phi=0,\dot{\phi}=0)\}$.
In summary, we have proven the following result: 
\begin{quote}
Up to normalization, there exists a unique conserved measure on the
effective phase space $(\phi,\dot{\phi})$ for scalar-field cosmologies
with $m^{2}\phi^{2}$ potentials, excluding the origin. 
\end{quote}
It is well-founded to conjecture that the effective phase-space measure
always exists for any reasonable potential $V\left(\phi\right)$.
A requisite property of the potential is that, if the the measure
diverges on a set contained in a neighborhood $W$ (e.g., during reheating),
there is a unique solution in an open neighborhood $U\backslash W$,
cf. Eq. \eqref{lateU}. This is equivalent to the statement that the
measure on $K$ does not behave chaotically as the boundary of $U$
with $W$ is varied.

\section{The Physical Meaning of Attractors\label{PhysicalMeaning}}

We have seen that it is possible to define a conserved measure on
$K$, the effective $\phi$-$\dot{\phi}$ phase space of scalar-field
cosmology in a flat universe. This was shown for $V(\phi)=m^{2}\phi^{2}/2$,
and seems likely to hold for other smooth potentials in single-field
models. Because Liouville's theorem is obeyed with respect to such
a measure (unlike the na\"{i}ve measure ${\rm d}\dot{\phi}\wedge{\rm d}\phi$),
true attractor behavior is impossible. The apparent attractor behavior
familiar in cosmology is an artifact of using the $\phi$-$\dot{\phi}$
coordinates, which are not canonically conjugate. It is nevertheless
worth asking whether there are other ways of thinking about attractors
that are physically meaningful. In this section we suggest two possibilities.

The first possibility rests on the idea that $\phi$ and $\dot{\phi}$,
while not canonically conjugate, are nevertheless the directly physically
observable features of the scalar field. In this sense they define
preferred coordinates in which to follow the evolution. If one accepts
that notion, we can define an apparent attractor as a region in phase
space for which Lyapunov exponents are highly negative along particular
axes. With trajectories $\mathbf{x}\left(t\right)$ in phase space
labeled by coordinates $x^{\alpha}$ indexed by $\alpha$, we define
the Lyapunov exponent \cite{Lyapunov} along each coordinate axis,
\begin{equation}
\lambda_{\alpha}=\lim_{t\rightarrow\infty}\lim_{\delta x^{\alpha}\left(0\right)\rightarrow0}\frac{1}{t}\log\frac{\left|\delta x^{\alpha}\left(t\right)\right|}{\left|\delta x^{\alpha}\left(0\right)\right|},
\end{equation}
where $\delta x^{\alpha}\left(t\right)$ is the separation between
two trajectories, in the $\alpha$-coordinate, at time $t$. (Note
that $\lambda_{\alpha}$ is a function of position in phase space.)
Then 
\begin{equation}
\left|\delta x^{\alpha}\left(t\right)\right|\sim e^{\lambda_{\alpha}t}\left|\delta x\left(0\right)\right|.
\end{equation}
By Liouville's theorem, the sum of the Lyapunov exponents in canonically
conjugate coordinates is zero and, in fact, a canonical transformation
of the coordinates on phase space can be made such that all the Lyapunov
exponents vanish \cite{ChienYao}. However, in the case of the linear
attractors for $m^{2}\phi^{2}$ potentials located near $\dot{\phi}=\pm\sqrt{2/3}m$,
the Lyapunov exponent in the $\dot{\phi}$ direction is negative,
forcing trajectories to appear (in $\phi$-$\dot{\phi}$ space) to
converge (cf. Fig \ref{attractors}). This definition of an apparent
attractor is consistent with the common motivation for using $\phi$-$\dot{\phi}$
space in the first place: the coordinates are physically intuitive
and trajectories exhibit ``attractor-like'' behavior. In contrast,
plotting in $(\phi,\pi_{\phi})$, where $\pi_{\phi}$ is the canonical
momentum \eqref{piphi} associated with $\phi$ under the Lagrangian
on effective phase space (Sec. \ref{momentum&measure}), will cause
the apparent attractor behavior to vanish. In $(\phi,\pi_{\phi})$
coordinates, bundles of trajectories will not shrink, but will instead
contract along one axis while expanding along the other. While the
Lyapunov exponent characterization is manifestly coordinate-dependent,
it has the virtues of being intuitive and capturing the sense in which
the word ``attractor'' is used in much of the literature, cf. Ref.
\cite{LyapAtr}.

Another point worth emphasizing is that our analysis here has been
entirely classical. In real cosmological evolution, there will be
a boundary in phase space past which a classical picture is inadequate;
we expect that this would occur at least when any physical quantity
(the Hubble constant, the radius of curvature, or the field energy)
reached the Planck scale. If one had a true theory of cosmological
initial conditions that implied a probability measure for trajectories
near this boundary, that would presumably supersede the classical
Liouville-type measures we have been focusing on in this paper. In
the absence of such a theory, of course, it makes more sense to use
such well-defined classical measures rather than to place too much
emphasis on the na\"{i}ve choice ${\rm d}\dot{\phi}\wedge{\rm d}\phi$.

\section{Conclusions}

It is common in literature on inflation (as well as quintessence models)
to make use of the idea of attractor solutions. However, this notion
is not well-defined for a Hamiltonian system. In this work, we have
attempted to clarify the relationship between the dissipationless
dynamics of scalar-field cosmology and apparent attractor behavior.

We showed that, in a universe with vanishing spatial curvature, there
is a sense in which $\phi$ and $\dot{\phi}$ become effective phase-space
variables. Namely, the map from the four-dimensional phase space to
$\phi$-$\dot{\phi}$ space is vector field invariant under the Hamiltonian
flow vector. As a result, trajectories do not cross in $\phi$-$\dot{\phi}$
space and, in this mapping, the phase space is effectively two-dimensional
and we observe the appearance of attractor-like behavior. In making
(coordinate-dependent) observations about ``attractors'', many authors
plot in $(\phi,\dot{\phi})$ coordinates, despite their noncanonicity,
and suppress the other two phase-space dimensions.

We explored the existence of a conserved measure on the effective
phase space. The GHS measure, while possessing many useful properties,
diverges for flat universes and so can give no information about the
measure on the effective phase space of $(\phi,\dot{\phi})$. Such
a measure can be constructed ``from scratch'' by finding a two-form
$\boldsymbol{\sigma}=f(\phi,\dot{\phi})\mathrm{d}\dot{\phi}\wedge\mathrm{d}\phi$
on $\phi$-$\dot{\phi}$ space that is conserved under Hamiltonian
flow (that is, whose Lie derivative along the Hamiltonian flow vector
vanishes). Using Douglas' theorem and the Helmholtz condition, we
proved that, for $V\left(\phi\right)$ inflation, such a measure constructed
in this way exists if and only if there is a Lagrangian description
of the system in the two-dimensional effective phase space. Furthermore,
using this Lagrangian, one can define a momentum conjugate to $\phi$
in the effective phase space, $\pi_{\phi}=\partial\mathcal{L}_{K}/\partial\dot{\phi}$
(not to be confused with $p_{\phi}$, the momentum conjugate to $\phi$
in the full four-dimensional phase space), and use this to define
a measure $\mathrm{d}\pi_{\phi}\wedge\mathrm{d}\phi$. We proved that
this measure is identical to the measure $f\,\mathrm{d}\dot{\phi}\wedge\mathrm{d}\phi$
that one constructs ``from scratch''; demanding conservation under
Hamiltonian flow is enough to specify the measure. For the specific
model of inflation with a quadratic potential, we found the behavior
that the effective phase-space measure must possess in the late ($H\ll m$)
and early ($H\gg m$) limits and used the Cauchy--Kowalevski theorem
to prove that a unique analytic solution for the measure exists up
to normalization, provided $\phi$ and $\dot{\phi}$ do not both vanish.
It is reasonable to conjecture that a similar existence/uniqueness
result should hold for a large class of potentials.

Finally, we discussed the meaning of apparent attractors. While the
dynamics of scalar-field cosmology is conservative, evolution can
nevertheless approach certain characteristics if we express it in
terms of preferred variables. It can happen, for example, that the
Lyapunov exponents can be negative along certain axes. By Liouville's
theorem, these more general formulations of apparent attractors are
necessarily coordinate-dependent.

The idea of attractor-like behavior is central to the intuitive idea
of inflationary cosmology: the development of a smooth, flat FRW universe
from a large set of initial conditions. Despite the fact that this
behavior cannot occur in canonical phase-space variables, it is useful
to consider how the notion of an apparent attractor can best be defined
to capture this intuition. This helps clarify the idea of naturalness
in cosmological evolution. 
\begin{acknowledgments}
We thank Alan Guth and Chien-Yao Tseng for helpful conversations.
Grant N. Remmen is supported by a Hertz Graduate Fellowship and an
NSF Graduate Research Fellowship. This material is based upon work
supported by the National Science Foundation Graduate Research Fellowship
Program under Grant No. DGE-1144469, by DOE grant DE-FG02-92ER40701,
and by the Gordon and Betty Moore Foundation through Grant 776 to
the Caltech Moore Center for Theoretical Cosmology and Physics. 
\end{acknowledgments}
 \bibliographystyle{apsrevGNR}
\bibliography{InflationBibliography}

\begin{thebibliography}{29}
\expandafter\ifx\csname natexlab\endcsname\relax\def\natexlab#1{#1}\fi
\expandafter\ifx\csname bibnamefont\endcsname\relax
  \def\bibnamefont#1{#1}\fi
\expandafter\ifx\csname bibfnamefont\endcsname\relax
  \def\bibfnamefont#1{#1}\fi
\expandafter\ifx\csname citenamefont\endcsname\relax
  \def\citenamefont#1{#1}\fi
\expandafter\ifx\csname url\endcsname\relax
  \def\url#1{\textrm{#1}}\fi
\expandafter\ifx\csname urlprefix\endcsname\relax\def\urlprefix{URL }\fi
\providecommand{\bibinfo}[2]{#2}
\providecommand{\eprint}[2][]{\url{#2}}

\bibitem[{\citenamefont{Guo \emph{et~al.}}(2003)\citenamefont{Guo, Piao, Cai,
  and Zhang}}]{tachyon}
\bibinfo{author}{\bibfnamefont{Z.-K.} \bibnamefont{Guo}},
  \bibinfo{author}{\bibfnamefont{Y.-S.} \bibnamefont{Piao}},
  \bibinfo{author}{\bibfnamefont{R.-G.} \bibnamefont{Cai}}, \bibnamefont{and}
  \bibinfo{author}{\bibfnamefont{Y.-Z.} \bibnamefont{Zhang}},
  \href{http://dx.doi.org/10.1103/PhysRevD.68.043508}{\bibinfo{journal}{Phys.
  Rev. D}}
  \href{http://dx.doi.org/10.1103/PhysRevD.68.043508}{\textbf{\bibinfo{volume}{68}},
  \bibinfo{pages}{043508}}\href{http://dx.doi.org/10.1103/PhysRevD.68.043508}{
  (\bibinfo{year}{2003})}.

\bibitem[{\citenamefont{Ure{\~n}a-L{\'o}pez and Reyes-Ibarra}(2009)}]{Urena}
\bibinfo{author}{\bibfnamefont{L.~A.} \bibnamefont{Ure{\~n}a-L{\'o}pez}}
  \bibnamefont{and} \bibinfo{author}{\bibfnamefont{M.~J.}
  \bibnamefont{Reyes-Ibarra}},
  \href{http://dx.doi.org/10.1142/S0218271809014674}{\bibinfo{journal}{Int. J.
  of Mod. Phys. D}}
  \href{http://dx.doi.org/10.1142/S0218271809014674}{\textbf{\bibinfo{volume}{18}},
  \bibinfo{pages}{621}}\href{http://dx.doi.org/10.1142/S0218271809014674}{
  (\bibinfo{year}{2009})}.

\bibitem[{\citenamefont{Belinsky \emph{et~al.}}(1985)\citenamefont{Belinsky,
  Grishchuk, Khalatnikov, and Zeldovich}}]{Belinsky}
\bibinfo{author}{\bibfnamefont{V.}~\bibnamefont{Belinsky}},
  \bibinfo{author}{\bibfnamefont{L.}~\bibnamefont{Grishchuk}},
  \bibinfo{author}{\bibfnamefont{I.}~\bibnamefont{Khalatnikov}},
  \bibnamefont{and}
  \bibinfo{author}{\bibfnamefont{Y.}~\bibnamefont{Zeldovich}},
  \href{http://dx.doi.org/10.1016/0370-2693(85)90644-6}{\bibinfo{journal}{Phys.
  Lett. B}}
  \href{http://dx.doi.org/10.1016/0370-2693(85)90644-6}{\textbf{\bibinfo{volume}{155}},
  \bibinfo{pages}{232 }}\href{http://dx.doi.org/10.1016/0370-2693(85)90644-6}{
  (\bibinfo{year}{1985})}.

\bibitem[{\citenamefont{Piran and Williams}(1985)}]{Piran}
\bibinfo{author}{\bibfnamefont{T.}~\bibnamefont{Piran}} \bibnamefont{and}
  \bibinfo{author}{\bibfnamefont{R.~M.} \bibnamefont{Williams}},
  \href{http://dx.doi.org/10.1016/0370-2693(85)90291-6}{\bibinfo{journal}{Phys.
  Lett. B}}
  \href{http://dx.doi.org/10.1016/0370-2693(85)90291-6}{\textbf{\bibinfo{volume}{163}},
  \bibinfo{pages}{331 }}\href{http://dx.doi.org/10.1016/0370-2693(85)90291-6}{
  (\bibinfo{year}{1985})}.

\bibitem[{\citenamefont{Ratra and Peebles}(1988)}]{Peebles}
\bibinfo{author}{\bibfnamefont{B.}~\bibnamefont{Ratra}} \bibnamefont{and}
  \bibinfo{author}{\bibfnamefont{P.~J.~E.} \bibnamefont{Peebles}},
  \href{http://dx.doi.org/10.1103/PhysRevD.37.3406}{\bibinfo{journal}{Phys.
  Rev. D}}
  \href{http://dx.doi.org/10.1103/PhysRevD.37.3406}{\textbf{\bibinfo{volume}{37}},
  \bibinfo{pages}{3406}}\href{http://dx.doi.org/10.1103/PhysRevD.37.3406}{
  (\bibinfo{year}{1988})}.

\bibitem[{\citenamefont{Liddle and Lyth}(2000)}]{LiddleLyth}
\bibinfo{author}{\bibfnamefont{A.}~\bibnamefont{Liddle}} \bibnamefont{and}
  \bibinfo{author}{\bibfnamefont{D.}~\bibnamefont{Lyth}},
  \emph{\bibinfo{title}{Cosmological Inflation and Large-Scale Structure}}
  (\bibinfo{publisher}{Cambridge University Press}, \bibinfo{year}{2000}).

\bibitem[{\citenamefont{Liddle \emph{et~al.}}(1994)\citenamefont{Liddle,
  Parsons, and Barrow}}]{Liddle}
\bibinfo{author}{\bibfnamefont{A.~R.} \bibnamefont{Liddle}},
  \bibinfo{author}{\bibfnamefont{P.}~\bibnamefont{Parsons}}, \bibnamefont{and}
  \bibinfo{author}{\bibfnamefont{J.~D.} \bibnamefont{Barrow}},
  \href{http://dx.doi.org/10.1103/PhysRevD.50.7222}{\bibinfo{journal}{Phys.
  Rev. D}}
  \href{http://dx.doi.org/10.1103/PhysRevD.50.7222}{\textbf{\bibinfo{volume}{50}},
  \bibinfo{pages}{7222}}\href{http://dx.doi.org/10.1103/PhysRevD.50.7222}{
  (\bibinfo{year}{1994})}.

\bibitem[{\citenamefont{Kiselev and Timofeev}(2008)}]{Kiselev}
\bibinfo{author}{\bibfnamefont{V.}~\bibnamefont{Kiselev}} \bibnamefont{and}
  \bibinfo{author}{\bibfnamefont{S.}~\bibnamefont{Timofeev}}
  (\bibinfo{year}{2008}), \href{http://arxiv.org/abs/0801.2453}{{\rmfamily
  arXiv:0801.2453 [gr-qc]}}.

\bibitem[{\citenamefont{Ferreira and Joyce}(1998)}]{Ferreira}
\bibinfo{author}{\bibfnamefont{P.~G.} \bibnamefont{Ferreira}} \bibnamefont{and}
  \bibinfo{author}{\bibfnamefont{M.}~\bibnamefont{Joyce}},
  \href{http://dx.doi.org/10.1103/PhysRevD.58.023503}{\bibinfo{journal}{Phys.
  Rev. D}}
  \href{http://dx.doi.org/10.1103/PhysRevD.58.023503}{\textbf{\bibinfo{volume}{58}},
  \bibinfo{pages}{023503}}\href{http://dx.doi.org/10.1103/PhysRevD.58.023503}{
  (\bibinfo{year}{1998})}.

\bibitem[{\citenamefont{Downes \emph{et~al.}}(2012)\citenamefont{Downes, Dutta,
  and Sinha}}]{Downes}
\bibinfo{author}{\bibfnamefont{S.}~\bibnamefont{Downes}},
  \bibinfo{author}{\bibfnamefont{B.}~\bibnamefont{Dutta}}, \bibnamefont{and}
  \bibinfo{author}{\bibfnamefont{K.}~\bibnamefont{Sinha}},
  \href{http://dx.doi.org/10.1103/PhysRevD.86.103509}{\bibinfo{journal}{Phys.
  Rev. D}}
  \href{http://dx.doi.org/10.1103/PhysRevD.86.103509}{\textbf{\bibinfo{volume}{86}},
  \bibinfo{pages}{103509}}\href{http://dx.doi.org/10.1103/PhysRevD.86.103509}{
  (\bibinfo{year}{2012})}.

\bibitem[{\citenamefont{Khoury and Steinhardt}(2011)}]{Steinhardt}
\bibinfo{author}{\bibfnamefont{J.}~\bibnamefont{Khoury}} \bibnamefont{and}
  \bibinfo{author}{\bibfnamefont{P.~J.} \bibnamefont{Steinhardt}},
  \href{http://dx.doi.org/10.1103/PhysRevD.83.123502}{\bibinfo{journal}{Phys.
  Rev. D}}
  \href{http://dx.doi.org/10.1103/PhysRevD.83.123502}{\textbf{\bibinfo{volume}{83}},
  \bibinfo{pages}{123502}}\href{http://dx.doi.org/10.1103/PhysRevD.83.123502}{
  (\bibinfo{year}{2011})}.

\bibitem[{\citenamefont{Clesse \emph{et~al.}}(2009)\citenamefont{Clesse,
  Ringeval, and Rocher}}]{LyapAtr}
\bibinfo{author}{\bibfnamefont{S.}~\bibnamefont{Clesse}},
  \bibinfo{author}{\bibfnamefont{C.}~\bibnamefont{Ringeval}}, \bibnamefont{and}
  \bibinfo{author}{\bibfnamefont{J.}~\bibnamefont{Rocher}},
  \href{http://dx.doi.org/10.1103/PhysRevD.80.123534}{\bibinfo{journal}{Phys.
  Rev. D}}
  \href{http://dx.doi.org/10.1103/PhysRevD.80.123534}{\textbf{\bibinfo{volume}{80}},
  \bibinfo{pages}{123534}}\href{http://dx.doi.org/10.1103/PhysRevD.80.123534}{
  (\bibinfo{year}{2009})}.

\bibitem[{\citenamefont{Gibbons \emph{et~al.}}(1987)\citenamefont{Gibbons,
  Hawking, and Stewart}}]{GHS}
\bibinfo{author}{\bibfnamefont{G.}~\bibnamefont{Gibbons}},
  \bibinfo{author}{\bibfnamefont{S.}~\bibnamefont{Hawking}}, \bibnamefont{and}
  \bibinfo{author}{\bibfnamefont{J.}~\bibnamefont{Stewart}},
  \href{http://dx.doi.org/10.1016/0550-3213(87)90425-1}{\bibinfo{journal}{Nucl.
  Phys. B}}
  \href{http://dx.doi.org/10.1016/0550-3213(87)90425-1}{\textbf{\bibinfo{volume}{281}},
  \bibinfo{pages}{736 }}\href{http://dx.doi.org/10.1016/0550-3213(87)90425-1}{
  (\bibinfo{year}{1987})}.

\bibitem[{\citenamefont{Milnor}(1985)}]{Milnor}
\bibinfo{author}{\bibfnamefont{J.}~\bibnamefont{Milnor}},
  \href{http://dx.doi.org/10.1007/BF01212280}{\bibinfo{journal}{Commun. Math.
  Phys.}}
  \href{http://dx.doi.org/10.1007/BF01212280}{\textbf{\bibinfo{volume}{99}},
  \bibinfo{pages}{177}}\href{http://dx.doi.org/10.1007/BF01212280}{
  (\bibinfo{year}{1985})}.

\bibitem[{\citenamefont{Auslander \emph{et~al.}}(1964)\citenamefont{Auslander,
  Bhatia, and Seibert}}]{Auslander}
\bibinfo{author}{\bibfnamefont{J.}~\bibnamefont{Auslander}},
  \bibinfo{author}{\bibfnamefont{N.}~\bibnamefont{Bhatia}}, \bibnamefont{and}
  \bibinfo{author}{\bibfnamefont{P.}~\bibnamefont{Seibert}},
  \href{http://zbmath.org/?q=an:0163.44802}{\bibinfo{journal}{Bol. Soc. Mat.
  Mexicana (2)}}
  \href{http://zbmath.org/?q=an:0163.44802}{\textbf{\bibinfo{volume}{9}},
  \bibinfo{pages}{55}}\href{http://zbmath.org/?q=an:0163.44802}{
  (\bibinfo{year}{1964})}.

\bibitem[{\citenamefont{Gibbs}(1902)}]{Gibbs}
\bibinfo{author}{\bibfnamefont{J.}~\bibnamefont{Gibbs}},
  \emph{\bibinfo{title}{Elementary Principles in Statistical Mechanics}}
  (\bibinfo{publisher}{Yale University Press}, \bibinfo{year}{1902}).

\bibitem[{\citenamefont{Misner \emph{et~al.}}(1973)\citenamefont{Misner,
  Thorne, and Wheeler}}]{MTW}
\bibinfo{author}{\bibfnamefont{C.}~\bibnamefont{Misner}},
  \bibinfo{author}{\bibfnamefont{K.}~\bibnamefont{Thorne}}, \bibnamefont{and}
  \bibinfo{author}{\bibfnamefont{J.}~\bibnamefont{Wheeler}},
  \emph{\bibinfo{title}{Gravitation}} (\bibinfo{publisher}{W. H. Freeman},
  \bibinfo{year}{1973}).

\bibitem[{\citenamefont{Tseng}(2013)}]{ChienYao}
\bibinfo{author}{\bibfnamefont{C.-Y.} \bibnamefont{Tseng}}, Ph.D. thesis,
  \bibinfo{school}{\href{http://resolver.caltech.edu/CaltechTHESIS:03212013-213920050}{California
  Institute of
  Technology}}\href{http://resolver.caltech.edu/CaltechTHESIS:03212013-213920050}{
  (\bibinfo{year}{2013})}.

\bibitem[{\citenamefont{Wald}(1984)}]{Wald}
\bibinfo{author}{\bibfnamefont{R.}~\bibnamefont{Wald}},
  \emph{\bibinfo{title}{General Relativity}} (\bibinfo{publisher}{University of
  Chicago Press}, \bibinfo{year}{1984}).

\bibitem[{\citenamefont{Carroll}(2004)}]{SeanGR}
\bibinfo{author}{\bibfnamefont{S.}~\bibnamefont{Carroll}},
  \emph{\bibinfo{title}{Spacetime and Geometry: An Introduction to General
  Relativity}} (\bibinfo{publisher}{Addison-Wesley Longman, Incorporated},
  \bibinfo{year}{2004}).

\bibitem[{\citenamefont{Carroll and Tam}(2010)}]{CarrollTam}
\bibinfo{author}{\bibfnamefont{S.~M.} \bibnamefont{Carroll}} \bibnamefont{and}
  \bibinfo{author}{\bibfnamefont{H.}~\bibnamefont{Tam}} (\bibinfo{year}{2010}),
  \href{http://arxiv.org/abs/1007.1417}{{\rmfamily arXiv:1007.1417 [hep-th]}}.

\bibitem[{\citenamefont{Gibbons and Turok}(2008)}]{GibbonsTurok}
\bibinfo{author}{\bibfnamefont{G.}~\bibnamefont{Gibbons}} \bibnamefont{and}
  \bibinfo{author}{\bibfnamefont{N.}~\bibnamefont{Turok}},
  \href{http://dx.doi.org/10.1103/PhysRevD.77.063516}{\bibinfo{journal}{Phys.
  Rev. D}}
  \href{http://dx.doi.org/10.1103/PhysRevD.77.063516}{\textbf{\bibinfo{volume}{77}},
  \bibinfo{pages}{063516}}\href{http://dx.doi.org/10.1103/PhysRevD.77.063516}{
  (\bibinfo{year}{2008})}.

\bibitem[{\citenamefont{Ade \emph{et~al.}}(2013)}]{Planck}
\bibinfo{author}{\bibfnamefont{P.}~\bibnamefont{Ade}}
  \bibnamefont{\emph{et~al.}} (\bibinfo{collaboration}{Planck Collaboration})
  (\bibinfo{year}{2013}), \href{http://arxiv.org/abs/1303.5076}{{\rmfamily
  arXiv:1303.5076 [astro-ph.CO]}}.

\bibitem[{\citenamefont{Schiffrin and Wald}(2012)}]{SchiffrinWald}
\bibinfo{author}{\bibfnamefont{J.~S.} \bibnamefont{Schiffrin}}
  \bibnamefont{and} \bibinfo{author}{\bibfnamefont{R.~M.} \bibnamefont{Wald}},
  \href{http://dx.doi.org/10.1103/PhysRevD.86.023521}{\bibinfo{journal}{Phys.
  Rev. D}}
  \href{http://dx.doi.org/10.1103/PhysRevD.86.023521}{\textbf{\bibinfo{volume}{86}},
  \bibinfo{pages}{023521}}\href{http://dx.doi.org/10.1103/PhysRevD.86.023521}{
  (\bibinfo{year}{2012})}.

\bibitem[{\citenamefont{Hawking and Page}(1988)}]{HawkingPage}
\bibinfo{author}{\bibfnamefont{S.}~\bibnamefont{Hawking}} \bibnamefont{and}
  \bibinfo{author}{\bibfnamefont{D.~N.} \bibnamefont{Page}},
  \href{http://dx.doi.org/10.1016/0550-3213(88)90008-9}{\bibinfo{journal}{Nucl.
  Phys. B}}
  \href{http://dx.doi.org/10.1016/0550-3213(88)90008-9}{\textbf{\bibinfo{volume}{298}},
  \bibinfo{pages}{789 }}\href{http://dx.doi.org/10.1016/0550-3213(88)90008-9}{
  (\bibinfo{year}{1988})}.

\bibitem[{\citenamefont{Douglas}(1941)}]{Douglas}
\bibinfo{author}{\bibfnamefont{J.}~\bibnamefont{Douglas}},
  \href{http://www.jstor.org/stable/1989912}{\bibinfo{journal}{Trans. Amer.
  Math. Soc.}}
  \href{http://www.jstor.org/stable/1989912}{\textbf{\bibinfo{volume}{50}},
  \bibinfo{pages}{71}}\href{http://www.jstor.org/stable/1989912}{
  (\bibinfo{year}{1941})}.

\bibitem[{\citenamefont{Santilli}(1978)}]{Santilli}
\bibinfo{author}{\bibfnamefont{R.}~\bibnamefont{Santilli}},
  \emph{\bibinfo{title}{Foundations of Theoretical Mechanics I: The Inverse
  Problem in Newtonian Mechanics}} (\bibinfo{publisher}{Springer-Verlag},
  \bibinfo{year}{1978}).

\bibitem[{\citenamefont{Kowalevski}(1875)}]{Kowalevski}
\bibinfo{author}{\bibfnamefont{S.}~\bibnamefont{Kowalevski}},
  \href{http://dx.doi.org/10.1515/crll.1875.80.1}{\bibinfo{journal}{J. Reine
  Angew. Math.}}
  \href{http://dx.doi.org/10.1515/crll.1875.80.1}{\textbf{\bibinfo{volume}{80}},
  \bibinfo{pages}{1}}\href{http://dx.doi.org/10.1515/crll.1875.80.1}{
  (\bibinfo{year}{1875})}.

\bibitem[{\citenamefont{Lyapunov}(1992, orig. 1892)}]{Lyapunov}
\bibinfo{author}{\bibfnamefont{A.}~\bibnamefont{Lyapunov}},
  \emph{\bibinfo{title}{The General Problem of the Stability of Motion}},
  Control Theory and Applications Series (\bibinfo{publisher}{Taylor \&
  Francis}, \bibinfo{year}{1992, orig. 1892}), \bibinfo{note}{translated by
  A.T. Fuller}.

\end{thebibliography}

\end{document}